\documentclass[preprint,12pt]{elsarticle} 
\usepackage{booktabs}
\usepackage{amssymb}
\usepackage[colorlinks]{hyperref}   
\usepackage{xcolor}
\usepackage{amsthm}
\usepackage{mathtools}
\usepackage{enumitem}
\usepackage{graphicx}
\usepackage{subfigure}
\usepackage{autobreak}
\usepackage{lineno}
\usepackage{natbib}
\usepackage[margin=2cm]{geometry}
\usepackage{multirow}
\usepackage{floatrow}
\usepackage{colortbl}
\usepackage{threeparttable}

\usepackage{graphicx}
\newfloatcommand{capbtabbox}{table}[][\FBwidth]
\usepackage{multirow}

\setcitestyle{authoryear,open={(},close={)}}
\definecolor{orange}{rgb}{1,0.5,0}
\definecolor{red}{RGB}{198,0,35}
\definecolor{amberseldef}{rgb}{1.0, 0.49, 0.0}
\definecolor{ceruleanblue}{rgb}{0.16, 0.32, 0.75}
\definecolor{amber}{rgb}{1.0, 0.49, 0.0}
\definecolor{dodgerblue}{rgb}{0.12, 0.56, 1.0}
\definecolor{pureblue}{rgb}{0, 0, 1.0}

\definecolor{blue}{rgb}{0.0, 0.28, 0.67}

\def\hmath$#1${\texorpdfstring{{\rmfamily\textit{#1}}}{#1}}

\makeatletter
\def\ps@pprintTitle{%
   \let\@oddhead\@empty
   \let\@evenhead\@empty
   \let\@oddfoot\@empty
   \let\@evenfoot\@oddfoot
}
\makeatother

\AtBeginDocument{\hypersetup{citecolor= pureblue,linkcolor = pureblue,urlcolor = dodgerblue}}
\begin{document}


\begin{frontmatter}


\address[1]{Department of Civil Engineering, the University of Hong Kong, Hong Kong, China}
\address[2]{College of Civil Engineering and Architecture, Zhejiang University, Hangzhou, China}
\cortext[cor1]{Corresponding author: kejintao@hku.hk}

\author[1]{Wang Chen}
\author[1]{Jintao Ke \texorpdfstring{\corref{cor1}}{}}
\author[2]{Xiqun (Michael) Chen}

\title{Quantifying traffic emission reductions and traffic congestion alleviation from high-capacity ride-sharing}

\begin{abstract}

Despite the promising benefits that ride-sharing offers, there has been a lack of research on the benefits of high-capacity ride-sharing services. Prior research has also overlooked the relationship between traffic volume and the degree of traffic congestion and emissions. To address these gaps, this study develops an open-source agent-based simulation platform and a heuristic algorithm to quantify the benefits of high-capacity ride-sharing with significantly lower computational costs. The simulation platform integrates a traffic emission model and a speed-density traffic flow model to characterize the interactions between traffic congestion levels and emissions. The experiment results demonstrate that ride-sharing with vehicle capacities of 2, 4, and 6 passengers can alleviate total traffic congestion by approximately 3\%, 4\%, and 5\%, and reduce traffic emissions of a ride-sourcing system by approximately 30\%, 45\%, and 50\%, respectively. This study can guide transportation network companies in designing and managing more efficient and environment-friendly mobility systems.
\end{abstract}


\end{frontmatter}


\newpage

\section{Introduction}\label{sec: 1}
With massive population growth, large-scale urbanization, and rapid technological development, diverse urban mobility demands have increased significantly, bringing tremendous pressure on urban traffic systems and inducing various environmental pollution issues\citep{mcdonnell2016ecological, batty2008size, bettencourt2013origins}. On the one hand, in the process of urban expansion over the past decades, urban congestion has become costly in terms of time, money, and fuel \citep{lu2021expansion, ewing2018does, li2019does, chang2017there}. For example, it is estimated that traffic congestion generated approximately 8.8 billion hours of total delay for 494 U.S. urban areas in 2017, requiring around 3.3 billion gallons of excess fuel consumption and 179 billion dollars of cost \citep{schrank2019urban}. In Europe, the cost of traffic congestion is approximately equivalent to 1\% of GDP \citep{christidis2012measuring}. In China, there have been more than 80 cities with at least 1 million cars, causing serious traffic congestion \citep{li2019does}. On the other hand, traffic emissions contribute substantially to primary particulate matter (PM) (Table \ref{table1} lists all symbols and abbreviations used in this study) emissions in urban areas \citep{pant2013estimation}, which results in an enormous challenge to the global ecological environment and human health \citep{grimm2008global, EEA2022trans}. In 2022, global carbon dioxide (CO$_2$) emissions by the transport sector are around 8 Gt, accounting for approximately 25\% of CO$_2$ emissions from fuel combustion \citep{IEA2023CO2}. In Europe, the transport sector is a significant contributor to the emissions of air pollutants, accounting for approximately 10\% of PM2.5 emissions and 34\% of total nitrogen oxides (NOx), which comprises mostly nitrogen monoxide (NO) and nitrogen dioxide (NO$_2$) \citep{EEA2022trans,mulholland2022role}. In China, it is estimated that emissions from mobility-related activities account for approximately 83\% of hydrocarbons (HC), 46\% of NOx, and 78\% of carbon monoxide (CO) \citep{wang2008road}.

Traditionally, urban mobility needs are primarily satisfied by mass public transportation (including metro and buses), taxis, and private cars \citep{vuchic2017urban}. In recent years, the proliferation of mobile internet and wireless communication technologies has led to the rapid expansion of transportation network companies (TNCs) that provide ride-sourcing services. As a result, TNCs have become a critical player in meeting the urban mobility needs of commuters, serving as a reliable alternative to traditional modes of transportation \citep{wang2019ridesourcing}. For example, Uber, an international ride-sourcing company providing various on-demand mobility services in more than 700 metropolitan areas, has finished approximately 21 million daily trips in 2022 \citep{DMR2023a}. Didi, the largest on-demand mobility platform in China, offers a variety of services to 587 million users in more than 400 cities in China, with approximately 27 million daily trips in 2019 \citep{DMR2023b}. Additionally, TNCs account for 15\% of all intra-San Francisco vehicle trips, which is 12 times that of taxi trips \citep{castiglione2016tncs}. However, the vehicle occupancy rate is generally low with on average less than 1.5 passengers in each vehicle \citep{li2021does, mitchell2010reinventing}, resulting in low transportation efficiency and traffic congestion \citep{erhardt2019transportation, wu2021assessing, schaller2021can}. Certainly, improving vehicle occupancy rates can help improve transportation efficiency and reduce the required fleet size, thereby decreasing traffic emissions and alleviating traffic congestion.

To enhance urban mobility efficiency and reduce environmental impacts, many TNCs, such as Didi, Uber, and Lyft, have launched ride-sharing services that allow a driver to transport multiple passengers with similar origins and destinations in a single trip.\citep{ke2020pricing}. Additionally, with the rapid development of mobile internet technology, dynamic ride-sharing services that match passengers and drivers on very short notice are gaining popularity worldwide \citep{shaheen2015shared, agatz2012optimization, chen2017understanding}. These dynamic ride-sharing services can improve vehicle occupancy rates while maintaining the flexibility of drivers \citep{mcdonnell2016ecological, un2015world, storch2021incentive}. Ride-sharing programs can not only bring numerous benefits to passengers and drivers, such as saving costs and increasing revenue, but also benefit society by improving vehicle occupancy rates, reducing environmental impacts, and alleviating traffic congestion \citep{chan2012ridesharing, furuhata2013ridesharing, santi2014quantifying}. For example, \cite{santi2014quantifying} proved that on-demand ride-sharing services in New York City could reduce 40\% or even more of cumulative vehicle miles traveled (VMT). In addition, \cite{anair2020ride} showed that ride-sharing services can reduce traffic emissions by 33\% compared with normal ride-sourcing services. Therefore, ride-sharing is a promising means of reducing air pollution and alleviating traffic congestion.

Although a number of studies have been directed towards estimating the extent to which traffic emissions can be reduced after ride-sharing services are adopted \citep{tikoudis2021ridesharing, sui2019gps, tikoudis2021ridesharing, yan2020quantifying, cai2019environmental}, there are still a few unsolved questions. First, quantifying the implication of high-capacity ride-sharing services (with three or more passengers sharing one vehicle) is still a challenging task, as solving high-capacity ride-sharing problems is significantly time-consuming. Second, the interplay between traffic flow and traffic speed is usually neglected in previous studies for the sake of simplicity, but its impact on traffic emission calculation is non-trivial \citep{tachet2017scaling, yan2020quantifying}. Fewer on-road vehicles can result in higher vehicle speeds, thereby improving transportation efficiency and leading to a greater reduction in traffic emissions. Therefore, it is imperative to account for traffic congestion when calculating traffic emissions. Third, there is a lack of open-source easy-operating simulation platforms for high-capacity ride-sharing systems for the calculation of traffic emissions and estimation of traffic congestion. This study seeks to address these gaps by developing an agent-based ride-sharing simulation platform and a heuristic algorithm that can efficiently solve high-capacity ride-sharing problems with high accuracy and significantly lower computational costs. Additionally, this study integrates a traffic emission model and a speed-density traffic flow model into the simulation platform to evaluate the impacts of high-capacity ride-sharing services on reducing traffic emissions, while taking into account the effects of traffic congestion. 

\begingroup
\setlength{\tabcolsep}{6pt} 
\renewcommand{\arraystretch}{1} 
\begin{table}[!htbp]
\caption{List of symbols and abbreviations used in this study.}
\begin{center}
\begin{tabular}{ c|c } 
\hline
 Symbol or abbreviation & Description \\
\hline
PM           &  Particulate matter \\
CO$_2$       &  Carbon dioxide  \\
NO           &  Nitrogen monoxide \\
NO$_2$       &  Nitrogen dioxide \\
NOx          &  Nitrogen oxides, comprising mostly NO and NO$_2$ \\
TNC          & Transportation network company \\
VMT          & Vehicle miles traveled \\
FC           & Fuel consumption \\
COPERT       & Computer programme to calculate emissions from road transport \\
$EF_{i,j,k}$ & Emission factor of pollutant $k$ emitted by vehicle $i$ on road $j$ (g/km) \\
$v_{i,j}$    & Average speed of vehicle $i$ on road $j$ (km/h) \\
$\alpha_{k}$, $\beta_{k}$, $\gamma_{k}$, $\delta_{k}$, $\epsilon_{k}$, $\zeta_{k}$, $\eta_{k}$ & Parameters for calculating pollutant $k$ \\
$E_{k}$      & Amount of pollutant $k$ emitted by all vehicles \\
$l_{j}$      & Length of road $j$ \\
$u$          & Traffic speed \\
$u_{0}$      & Free-flow speed \\
$k$          & Traffic density \\
$k_{j}$      & Traffic jam density \\
$k_b$        & Basic traffic density \\
$k_r$        & Ride-sourcing density \\
ILP          & Integer linear programming \\
NN           & Nearest neighbor algorithm \\
$n$          & Number of served passengers \\
$N$          & Total number of passengers \\
PMD          & Passenger miles delivered \\
SR           & Service rate \\
PEF          & Passenger emission factor \\
$PEF_k$      & PEF of Pollutant $k$ \\
DF           & Delay factor \\
$DF_{t,j}$   & Delay factor for each road $j$ at observation time point $t$  \\ 
$u_{t,j}$    & Average space speed of road $j$ at time point $t$ \\
$T$          & Number of observation time points \\
$J$          & Number of roads \\

\hline
\end{tabular}
\label{table1}
\end{center}
\end{table}

\section{Literature review}\label{sec: 1.5}

This section presents a summary of existing studies regarding how ride-sharing services impact traffic emissions and traffic congestion, and the corresponding methods for quantifying the effects of ride-sharing. Some studies focus more on the impacts of ride-sharing services on traffic emissions, while others emphasize more on their influences on traffic congestion.

Regarding traffic emission reductions, the typical approach in previous studies is to utilize GPS trajectory data to estimate the fuel consumption and carbon emissions of both taxis and ride-sourcing vehicles \citep{weng2017taxi, sun2018analyzing, sui2019gps}. This methodology allows for a thorough analysis of the carbon footprint of these transportation modes and provides insights into potential strategies for reducing emissions. A few recent works quantified the environmental benefits of ride-sharing services by combining GPS trajectories and historical trip requests \citep{liu2021quantifying, cai2019environmental, zhu2022potential, zhang2020mobile}. In these studies, all shareable trips were initially identified based on predefined constraints for the origins, destinations, departure time, and arrival time of two passengers. The cumulative VMT were then calculated under a hypothetical scenario in which ride-sharing is adopted. The calculated results were compared with those under the original scenario with non-sharing ride-sourcing services. Based on this comparison, the reduction in VMT and traffic emissions resulting from ride-sharing can be calculated. In addition, a spatiotemporal analysis of emission reductions can be done to indicate at which time period and location ride-sharing can reduce more carbon emissions \citep{li2021does}. Moreover, other studies have explored the impacts of various travel-related and built environmental factors on carbon emission reductions resulting from ride-sharing, thereby identifying the most critical factors that can enhance the benefits of ride-sharing \citep{li2022revealing, yin2018appraising}. For example, \cite{li2022revealing} proved that the overlap rate and detour rate of shared rides are the most critical determinants for enhancing the traffic emission reduction rate of ride-sharing. Additionally, a few studies quantified the impacts of ride-sharing associated with autonomous or electric vehicles on traffic emissions \citep{morfeldt2022impacts, akimoto2022impacts}. For instance, \cite{morfeldt2022impacts} claimed that shared mobility could decrease the carbon footprint by approximately 41\% by 2050 if one shared vehicle replaces ten individual vehicles. Some researchers also used statistics from questionnaires filled by ride-sharing passengers to calibrate the calculation results \citep{chen2021exploring, si2022can}. Interestingly, \cite{si2022can} found that carbon-emission reduction certification can significantly increase users' willingness to participate in ride-sharing programs. However, most of the aforementioned studies assumed a ride-sharing scenario in which at most two passengers share one vehicle, allowing all possible sharing trips to be easily enumerated. For example, a driver would sequentially visit passenger A's origin, passenger B's origin, passenger A's destination, and passenger B's destination. However, this method cannot be extended to high-capacity ride-sharing scenarios due to the exponential increase in computational complexity with vehicle capacity. Therefore, the key question for high-capacity ride-sharing problems is how to accelerate the calculation process. This study addresses this challenge by developing a heuristic algorithm that achieves over 95\% accuracy with no more than 2\% computational costs compared with the enumeration method.

Regarding ride-sharing's implications for traffic congestion, there is common sense that ride-sharing can reduce the required vehicle fleet size to meet a given level of passenger demand, thereby alleviating traffic congestion, particularly in areas with high-density demand during peak hours \citep{shaheen2018shared, alisoltani2021can, zhang2022mitigating, engelhardt2019quantifying}. Specifically, \cite{alisoltani2021can} employed a dynamic trip-based macroscopic simulation to measure the congestion effect and dynamic travel times, finding that ride-sharing can significantly alleviate traffic congestion, especially in areas with high-density demand. \cite{zhang2022mitigating} demonstrated that encouraging ride-sharing through a policy that charges trips with only one passenger starting or ending in the urban center can reduce traffic congestion modestly and improve social welfare substantially. \cite{engelhardt2019quantifying} claimed that ride-sharing can benefit the entire transportation system, particularly by reducing traffic congestion on major roads when the passenger adoption rate for ride-sharing exceeds 5\%. In addition, \cite{ke2020ride} found that ride-sharing can alleviate traffic congestion when demand density is high, as vehicles in a ride-sharing market can run faster than those in a non-sharing ride-sourcing market. \cite{zhang2022ride} showed that the system efficiency can decrease due to the congestion effect, and increasing sharing size (i.e., the maximal vehicle capacity) can improve the system efficiency in many cases. \cite{correa2019congestion} proposed a congestion-aware ride-sharing algorithm that optimizes the travel plans for all ride-sharing trips within a specific time interval, which can alleviate traffic congestion. Although many studies have demonstrated that ride-sharing can alleviate traffic congestion as well as improve transportation efficiency from various perspectives, few studies have considered these effects when calculating traffic emissions. To address this gap, this study integrates a speed-density traffic flow model with a traffic emission model, enabling our developed simulation platform to account for traffic congestion when estimating traffic emissions.

Several quantifying and simulation methods have been proposed to address ride-sharing tasks. For instance, \cite{santi2014quantifying} proposed the shareability network, which represents each trip as a node, each route that serves two shareable trips via ride-sharing as a link, and the weight of the shared trip as the link weight, to solve the ride-sharing assignment problem. This approach has been widely applied in various ride-sharing research areas \citep{yan2020quantifying, guo2022shareability, koutsopoulos2023increasing}. For example, \cite{yan2020quantifying} adopted the shareability network and a speed-density traffic-flow model to quantify the benefits of ride-sharing in Shanghai, concluding that ride-sharing can reduce fuel consumption (FC) by 22.88\% and 15.09\% in optimal and realistic scenarios, respectively. Moreover, \cite{alonso2017demand} proposed the high-capacity ride-sharing strategy that collects all trip requests for a batch-matching time frame and matches all requests and vehicles. This can increase the efficiency of high-capacity ride-sharing systems, as systems gain more knowledge from all possible assignments. This strategy has been applied to various ride-sharing optimization approaches, such as reinforcement learning-based optimization \citep{shah2020neural}, stochastic optimization \citep{luo2021efficient}, and scaling laws analysis \citep{chen2023scaling}. Additionally, \cite{kucharski2020exact} developed the ExMAS algorithm, which can incrementally match single trips into shared trips based on passengers' utilities. This algorithm has been used to solve probabilistic ride-sharing and other problems \citep{soza2022shareability, kucharski2021modelling, bujak2022exploring}. Another widely used simulation tool for ride-sharing is the MATSim, a queuing-based network simulator \citep{w2016multi}, which has been extended and applied to various ride-sharing research areas, such as autonomous taxi services \citep{ruch2018amodeus, horl2017agent}, traffic noise simulation \citep{kuehnel2019noise, kuehnel2020impact}, and design and analysis of ride-sharing services \citep{zwick2021agent, zwick2020impact, ruch2020quantifying, bischoff2017city}. However, most of the existing simulation tools cannot address high-capacity ride-sharing problems or consider the impact of vehicle speed changes on transportation systems. To address this gap, this study develops an open-source agent-based ride-sharing simulator that takes into account traffic speeds and can quantify the effects of high-capacity ride-sharing services on reducing traffic emissions and alleviating traffic congestion.

\section{Data and simulation settings}\label{sec: 3}

\subsection{Data}\label{sec: 3.1}

The dataset used in this study comprises passenger requests provided by Didi Chuxing in Chengdu from November 1st to 30th, 2016. Each request includes an ID, timestamps, and origin and destination coordinates (longitude and latitude). The dataset contains approximately 200,000 requests per day in the whole city. To filter the dataset, we first select the urban zone that includes the Ring Expressway as the study area since most requests are generated in this area. Only requests with origins and destinations within the study area are considered. Second, we relocate all requests' origins and destinations to the nearest road intersections since this study assumes that vehicles can only pick up and drop off passengers at road intersections. Finally, since this study uses the batch matching algorithm, where passengers and vehicles are matched at regular time intervals, we reschedule the generation timestamps of all requests to the nearest batch-matching time points (see Section \ref{sec: 3.5} for details).

After filtering the data as described above, the dataset contains approximately 170,000 requests per day, and the temporal and distance distributions of the remaining requests are presented in Fig. \ref{fig: TemDis}. Notably, the temporal and distance distributions of requests are similar for all weekdays; therefore, this study selects requests from a typical day for simulation and analysis. As shown in Fig. \ref{fig: TemDis}(a), the request arrival rate is significantly higher during the daytime compared to that at midnight. Moreover, the average distance of requests is 6.4 km, and most requests have distances of no more than 15 km (Fig. \ref{fig: TemDis}(b)).

\begin{figure}[!htbp]
    \centering
    \subfigure[Temporal distribution]{\includegraphics[width=0.45\linewidth]{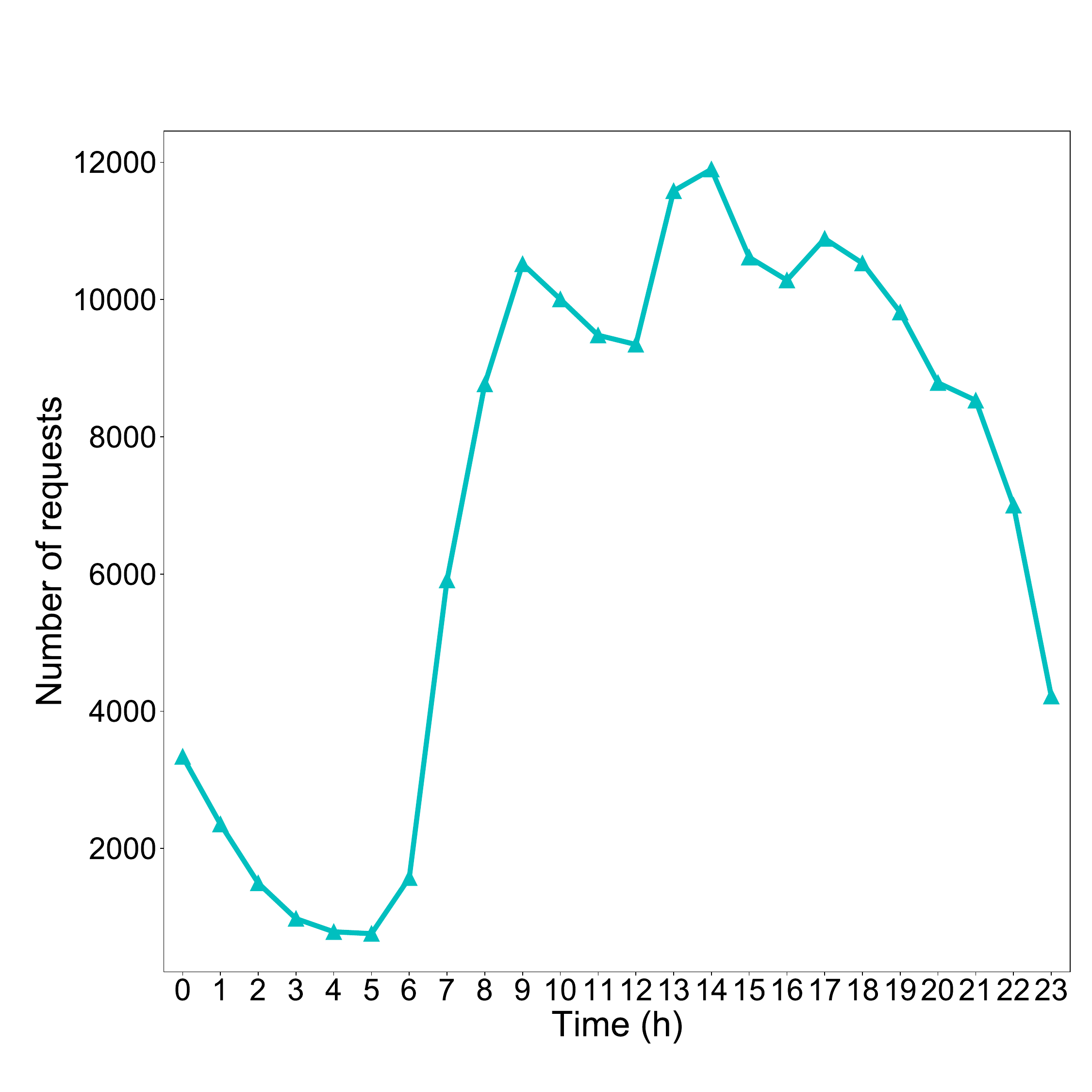}}
    \subfigure[Distance distribution]{\includegraphics[width=0.45\linewidth]{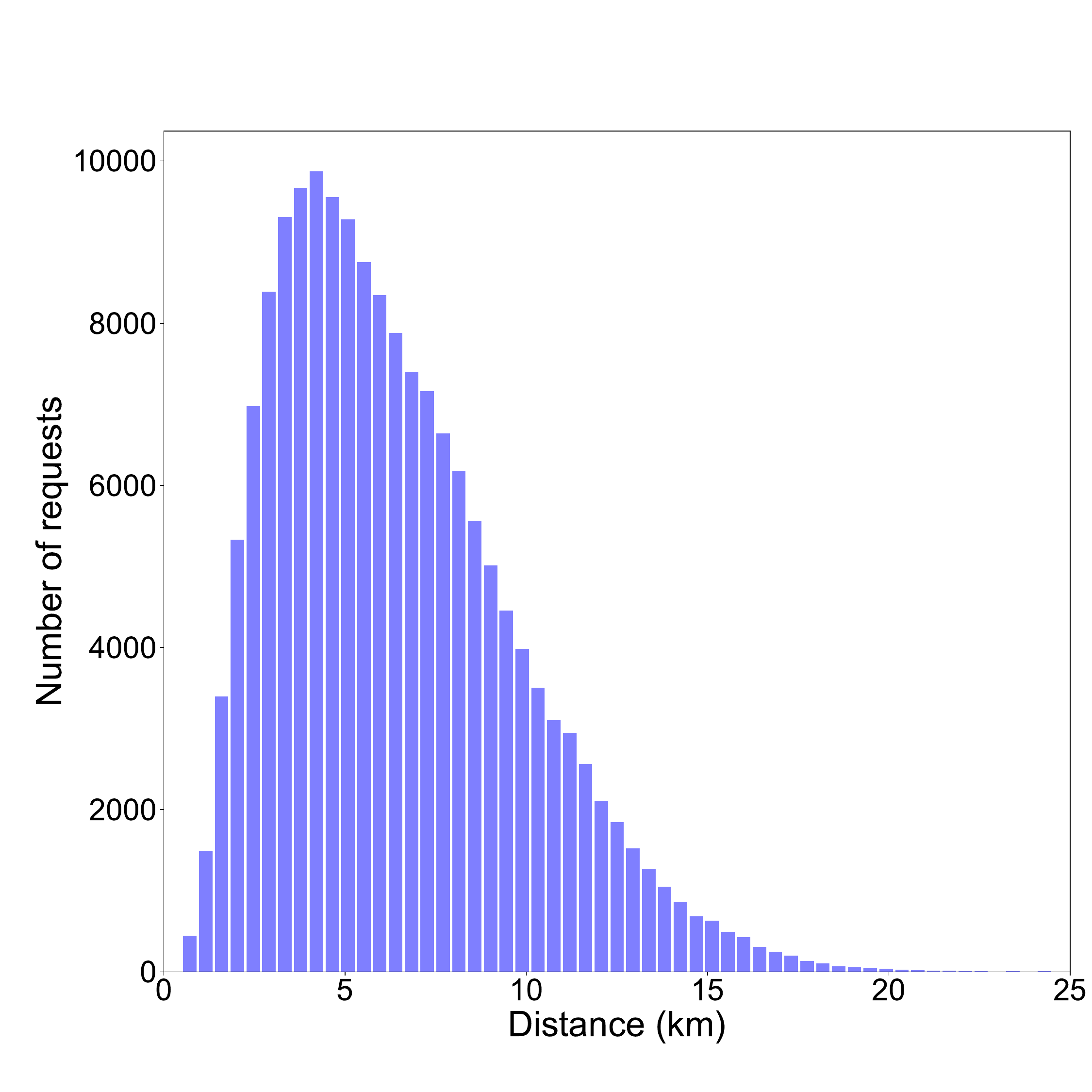}}
    \caption{Temporal and distance distributions of passenger requests on a typical day.}
    \label{fig: TemDis}
\end{figure}

To simulate the actions of vehicles and passengers on a real road network, this study uses the road network, including roads and intersections, obtained from Open Street Map \citep{OpenStreetMap}. The road network in the study area is relatively regular, as shown in Fig. \ref{fig: SpaDis}, ensuring that the simulation results in this area are typical since most cities worldwide are planned to be regular \citep{mumford1961city}. The majority of requests are concentrated in the northeastern zone of the study area, as shown in Fig. \ref{fig: SpaDis}(a). This study initializes vehicles according to the spatial distribution of requests; thus, most of the vehicles are initialized in the northeastern zone of the study area, as shown in Fig. \ref{fig: SpaDis}(b).

\begin{figure}[!htbp]
    \centering
    \subfigure[]{\includegraphics[width=0.55\linewidth]{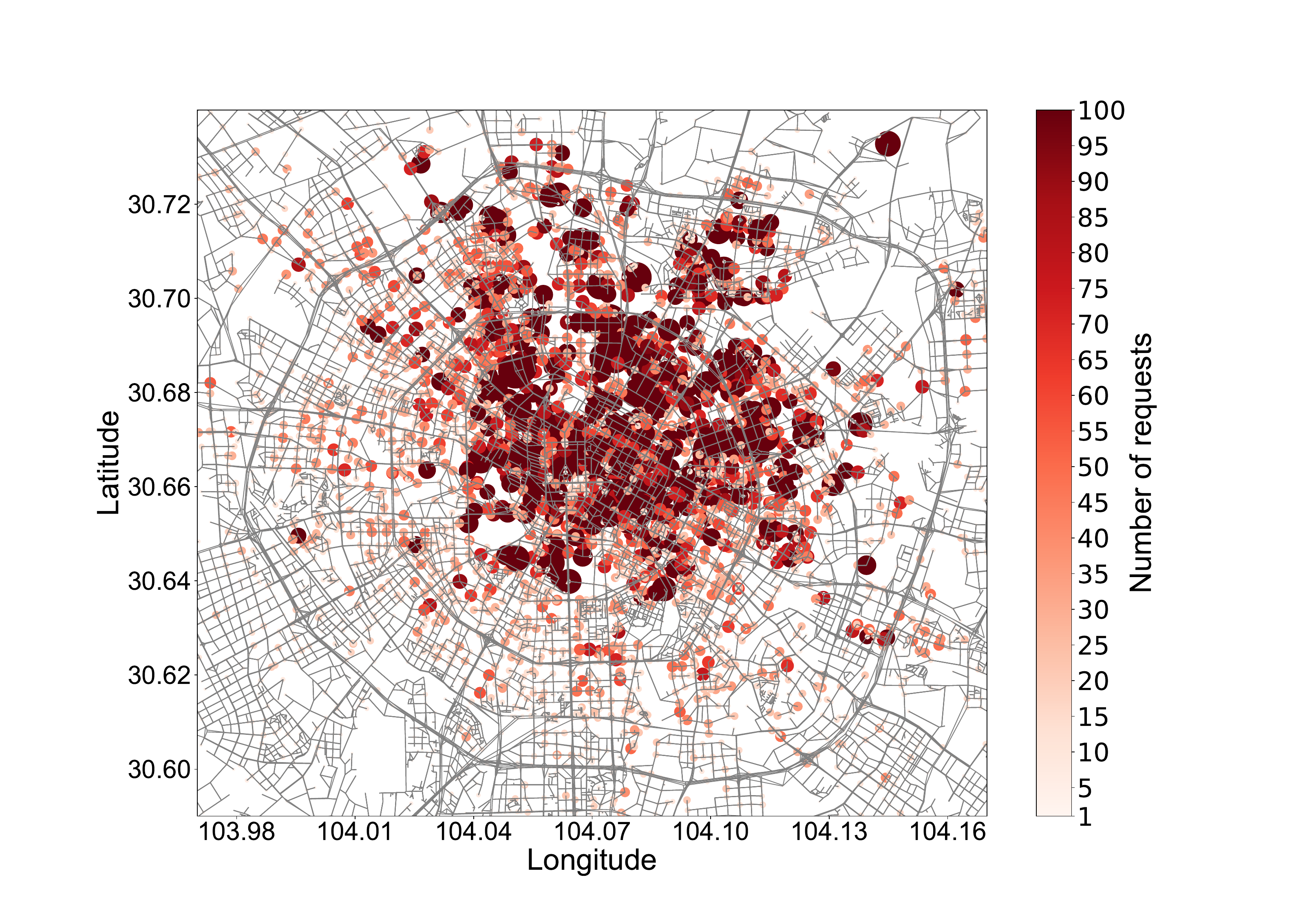}}
    \subfigure[]{\includegraphics[width=0.44\linewidth]{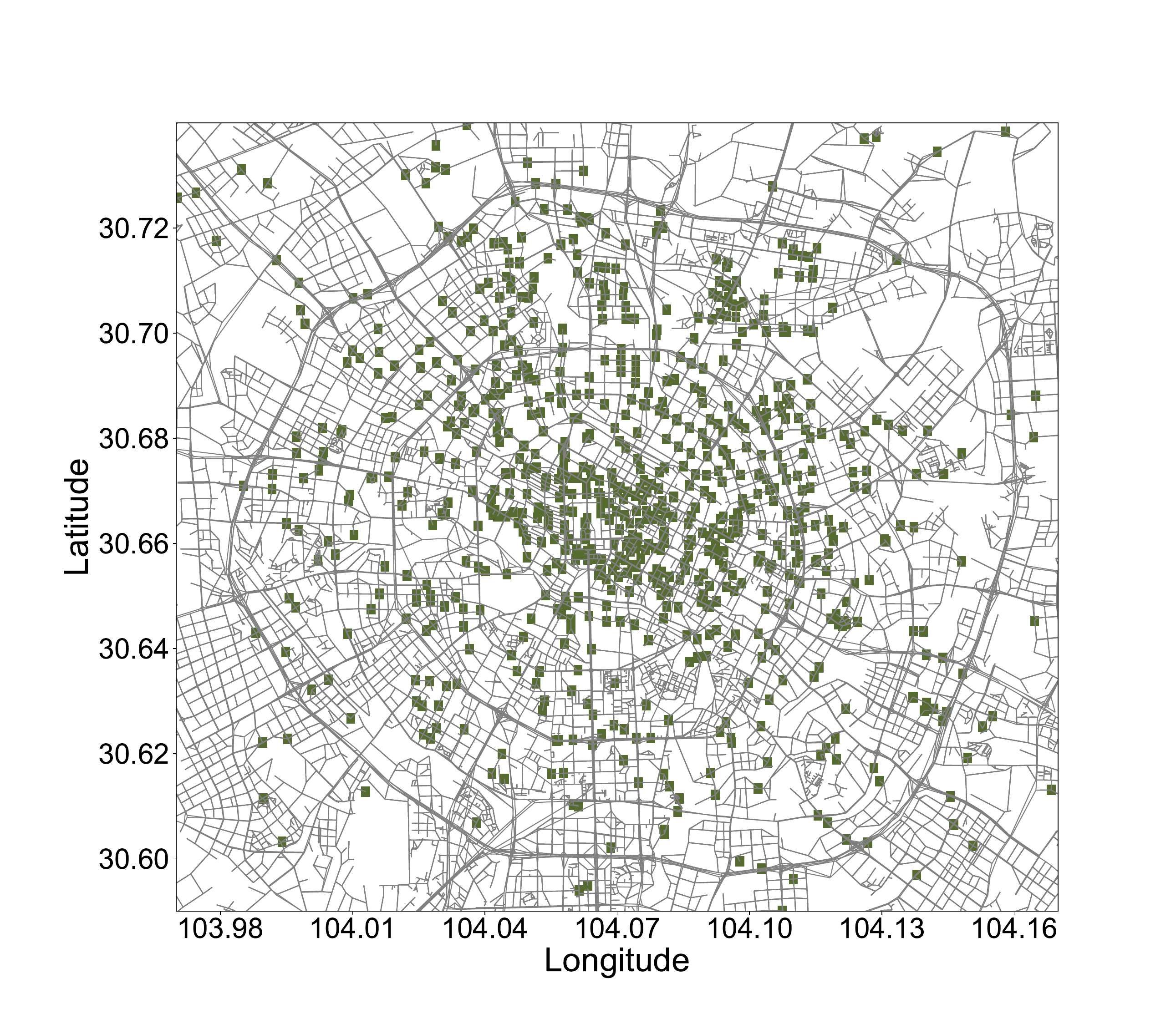}}
    \caption{Spatial distributions of (a) passenger requests and (b) vehicles. Red circles represent passengers in (a) and green squares represent vehicles in (b). The request arrival rate in (a) is 1.96 \#/s, and the number of vehicles in (b) is 1000.}
    \label{fig: SpaDis}
\end{figure}


\subsection{Simulation framework}\label{sec: 3.2}

We develop a ride-sharing simulation platform using Python \footnote{The code can be available at: \url{https://github.com/HKU-Smart-Mobility-Lab/Ride-sharing-Simulator}} which integrates the introduced heuristic algorithm (Sec. \ref{sec: 3.3}), traffic emission model (Sec. \ref{sec: 3.4}), and speed-density traffic flow model (Sec. \ref{sec: 3.5}). This simulation platform is developed based on the algorithm proposed by \cite{alonso2017demand}. However, the main differences between our simulation platform and the one proposed by \cite{alonso2017demand} are as follows. 

\begin{itemize}
    \item [(1)]
    This study develops a heuristic algorithm and verifies its effectiveness and efficiency for solving high-capacity ride-sharing problems, enabling the simulation platform to simulate high-capacity ride-sharing scenarios in a much more efficient way. This is important as researchers and practitioners may need to quantify the implications of high-capacity ride-pooling services in a large variety of market scenarios with varying supply and demand levels and city structures. 
    \item [(2)]
    This study integrates a traffic emission model and a speed-density traffic flow model into the simulation platform, enabling the platform to quantify the impact of ride-sharing services on reducing traffic emissions while considering traffic congestion. In particular, the simulation platform can dynamically update the shortest paths for vehicles according to real-time traffic congestion situations.
\end{itemize}

\begin{figure}[!htbp]
    \centering
    \includegraphics[width=0.9\linewidth]{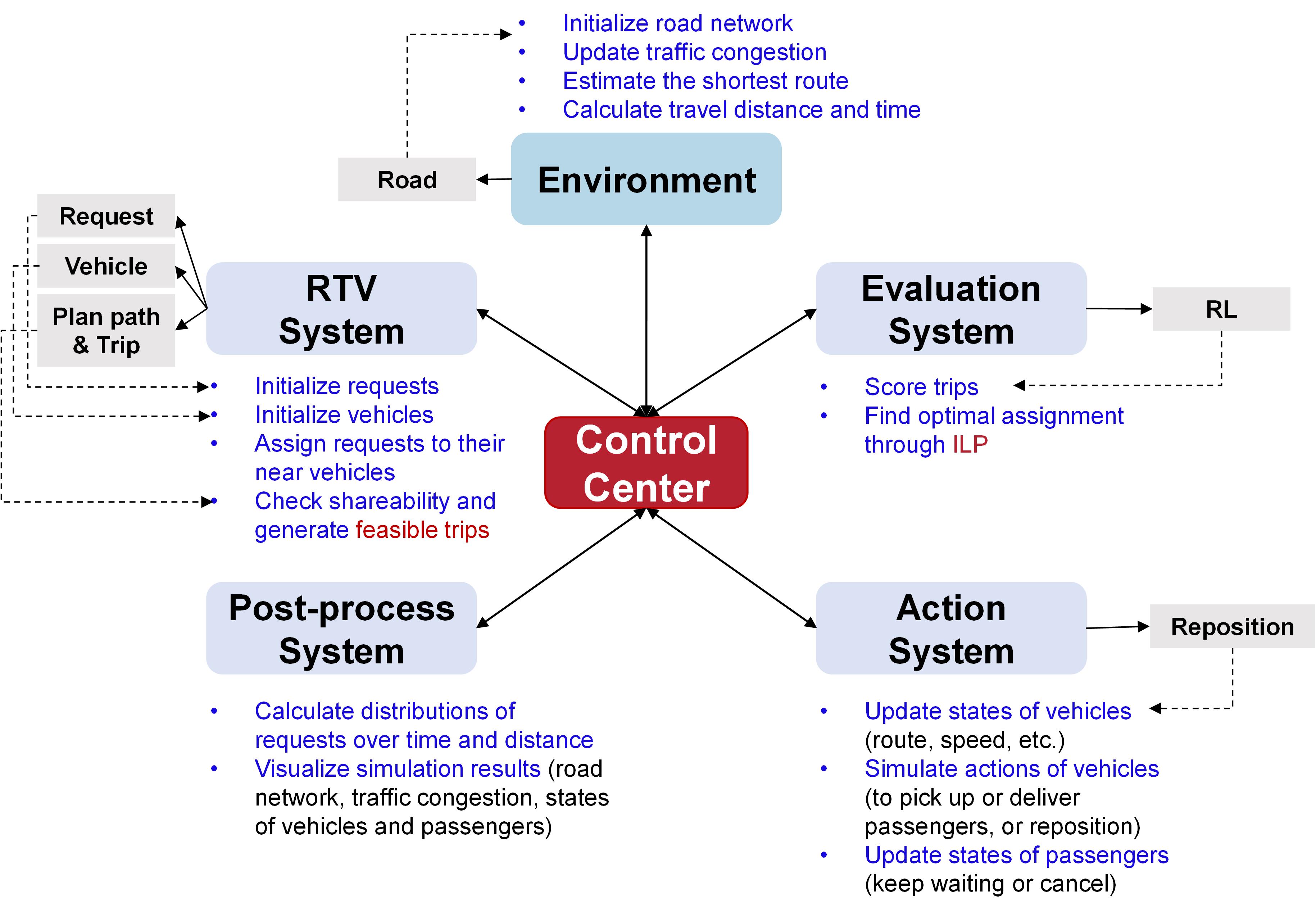}
    \caption{Architecture of the developed open-source agent-based ride-sharing simulation platform.}
    \label{fig: platform}
\end{figure}

The architecture of the simulation platform is shown in Fig. \ref{fig: platform}. The platform mainly consists of six parts, i.e., control center, environment, RTV system, evaluation system, action system, and post-process system. Control center is designed to integrate all functions, while environment initializes road networks and updates real-time traffic situations. In RTV system, each vehicle is regarded as an agent with its own attributes including the current on-vehicle passengers, the planned route, and so on. Agents drive to pick up and deliver their scheduled passengers according to the assignment results given by the integer linear programming (ILP) in evaluation system. Subsequently, agents' movements and speeds, and passengers' states are updated in action system. Finally, all simulation results are visualized in post-process system. \footnote{All parameters needed for the simulation platform, such as the maximum allowable detour time ratio and the vehicle capacity, are defined in a configuration file, enabling users to simulate various scenarios without needing to modify the source code.}


\subsection{Dispatching algorithm}\label{sec: 3.3}

The dispatching algorithm used in this study is similar to the one proposed by \cite{alonso2017demand}. Specifically, first, all passenger requests are collected within a batch-matching time interval (e.g., 2 seconds). Each request is potentially assigned to the vehicles within their matching area, i.e., the pickup distance is no more than the matching radius. Then, for each vehicle, the shareability of their potential request(s) and current request(s) is calculated based on the constraints of maximal pickup and detour time. Specifically, an optimal route with minimal costs for the vehicle to sequentially visit the origins and destinations of two passengers can be obtained by enumerating all possible routes and comparing their corresponding costs. If the optimal route meets all constraints of maximal pickup and detour time, the two passengers can share the vehicle, and are regarded as a feasible trip. Therefore, each vehicle might be preassigned with several feasible trips and each feasible trip is assigned a value according to the system objective. For example, the value can be the total price of the trip if the system objective is to maximize the revenue. It should be noted that each feasible trip contains one or more passenger requests and two different feasible trips may contain one same request. Subsequently, all vehicles and their feasible trips are matched through integer linear programming (ILP) that maximizes the total values of all feasible trips. Finally, vehicles pick up and deliver the assigned trips according to the matching results. Those passengers that have not been matched successfully at the current step will keep waiting for the next batch-matching frame until they are assigned to vehicles or quit the system due to the constraint of maximal waiting time. In addition, those idle vehicles that have not been assigned any passengers will cruise within near areas. Please refer to \cite{alonso2017demand} for details of the dispatching algorithm.

Under high-capacity scenarios, however, it is extremely time-consuming to enumerate all possible routes due to the tremendous number of possible routes \citep{santi2014quantifying, alonso2017demand}. Therefore, this study adopts the nearest neighbor (NN) algorithm to check the shareability among multiple requests. In particular, the NN algorithm guides the vehicle to visit the nearest spot (origin or destination) to pick up or drop off passengers. Thus, the locally optimal route through the NN algorithm can be obtained, which generally has slightly higher costs than the optimal route \citep{rosenkrantz1977analysis}.

To figure out the computational cost and accuracy of the NN algorithm, we conduct a numerical study in Chengdu, the same study area in Section \ref{sec: 3.1}, in which the fleet size is 50, the request arrival rate is around 170 requests per hour, and the time horizon is from 6 to 10 am. We compare the computational costs and accuracy of the NN algorithm with the enumeration method for two different scenarios with vehicle capacities of 4 and 6 passengers, respectively. The computational cost is represented as the total time for calculating the shareability among multiple requests during the whole simulation process, while the accuracy is calculated based on the confusion matrix, in which the calculation results of the NN algorithm and numeration method are respectively regarded as the predicted value and true label, as follows:

\begin{equation}
    \label{eq: acc}
    accuracy = \frac{TN+TP}{TN+FN+TP+FP},
\end{equation}
where $TP$ represents True Positive, i.e., the calculation results of both the NN algorithm and enumeration method are true. In other words, the multiple requests can be pooled to the driver. And $TN$, $FN$, and $FP$ represent True Negative, False Negative, and False Positive, respectively. Moreover, we define the scenario in which the locally optimal route given by the NN algorithm is the same as the optimal route given by the enumeration method as the same positive (SP). And we use $consistency$ to measure the ratio of SP to TP:

\begin{equation}
    \label{eq: consistency}
    consistency = \frac{SP}{TP}.
\end{equation}
This metric essentially measures the gap between the locally optimal route given by the NN algorithm and the optimal route given by the enumeration method.

The results of the numerical study are listed in Table \ref{table2}. The NN algorithm achieves $97.1\%$ accuracy and $91.6\%$ consistency with only $1.7\%$ computational costs, as well as $97.9\%$ accuracy and $92.3\%$ consistency with only $0.2\%$ computational costs in two scenarios with vehicle capacities of 4 and 6 passengers, respectively. It is worth noting that the accuracy and consistency in the scenario with higher vehicle capacity are slightly higher than that in the scenario with lower vehicle capacity. This is because there are more sharing possibilities among multiple requests in the scenario with higher vehicle capacity, as each vehicle can be preassigned with more requests. But actually, the generated feasible trips may be fewer since it is significantly more difficult to meet constraints for more requests. Therefore, there are more negative labels (a subgroup of passengers can not share a vehicle) in the scenario with higher vehicle capacity. Since it is easier to predict the negative values, the TN value in the scenario with higher vehicle capacity is significantly larger, making the accuracy slightly higher. In addition, the possible routes for higher-capacity ride-sharing is significantly less due to more constraints for more requests. Therefore, the NN algorithm has a greater possibility of providing the optimal route, making the value of consistency larger. The experimental results of the numerical study demonstrate that the NN algorithm can realize high accuracy with significantly lower computational costs compared with the enumeration method. This proves the NN algorithm is capable to be applied to high-capacity ride-sharing simulations.

\begingroup
\setlength{\tabcolsep}{10pt} 
\renewcommand{\arraystretch}{1} 

\begin{table}[!htbp]
\caption{Comparison between NN algorithm and enumeration method on computational costs and accuracy.}
\begin{center}
\begin{tabular}{ c|c|c|c|c }
\hline
& \multicolumn{2}{c}{Maximal vehicle capacity = 4}& \multicolumn{2}{|c}{Maximal vehicle capacity = 6} \\
\hline
&  Nearest Neighbor  & Enumeration & Nearest Neighbor   & Enumeration \\
\hline

Time (s)  &   8  &    483      & 12   &    5996     \\ 
Accuracy (\%)          & 97.1 &    --       & 97.9 &     --      \\
Consistency (\%)       & 91.6 &    --       & 92.3 &     --      \\

\hline
\end{tabular}
\label{table2}
\end{center}
\end{table}


\subsection{Emission estimation model}\label{sec: 3.4}

In this study, a widely used traffic emission estimation model developed by Europe Environment Agency, COPERT (i.e., COmputer Programme to calculate the Emissions from Road Transport) \citep{COPERT2020}, is adopted to calculate emissions. The COPERT model can estimate various pollutant levels emitted by different categories of vehicles using different types of fuels, such as carbon dioxide (CO$_2$), carbon monoxide (CO), nitrogen oxides (NO$_x$), hydrocarbon (HC), and so on. Moreover, when calculating emissions of various pollutants, the COPERT model takes into account different engine temperatures, driving situations, and climatic conditions. However, for simplicity, this study only focuses on the hot emissions (the engine is at its normal temperature) with urban driving. Since the COPERT model is developed based on Europe emission standard that is comparable to that implemented in China, a lot of research have proved the effectiveness of adopting the COPERT model to estimate vehicle emissions in Chinese cities \citep{cai2007estimation, lang2014air, luo2017analysis, sun2018analyzing, sui2019gps, li2021does}.

Given a specific category of vehicle and fuel, the COPERT model estimates the emission factor of various pollutants with respect to travel speed as follows:

\begin{equation}
    \label{eq: emission factor}
    EF_{i,j,k} = (\alpha_{k}v_{i,j}^2 + \beta_{k}v_{i,j} + \gamma_{k} + \delta_{k}/v_{i,j}) / (\epsilon_{k}v_{i,j}^2 + \zeta_{k}v_{i,j} + \eta_{k}),
\end{equation}
where $EF_{i,j,k}$ denotes the emission factor of pollutant $k$ emitted by vehicle $i$ on road $j$ (unit: g/km), $v_{i,j}$ the average speed of vehicle $i$ on road $j$ (unit: km/h), and $\alpha_{k}$, $\beta_{k}$, $\gamma_{k}$, $\delta_{k}$, $\epsilon_{k}$, $\zeta_{k}$, and $\eta_{k}$ are parameters for pollutant $k$, and their values can be found in \cite{COPERT2020}. Therefore, the amount of pollutant $k$ emitted by all vehicles $E_{k}$ can be calculated as:

\begin{equation}
    \label{eq: emissions of k}
    E_{k} = \sum_{i}{\sum_{j}{EF_{i,j,k}l_{j}}},
\end{equation}
where $l_{j}$ is the length of road $j$.

According to references \citep{sui2019gps, li2021does}, most ride-sourcing vehicles in Chengdu comply with the emission standard of China IV which is equal to Euro 4. Without loss of generality, this study assumes that all vehicles are Small Passenger Cars fueled with Petrol complying emission standard of Euro 4. As a result, the emission parameters of pollutant $k$ can be uniquely determined. Furthermore, the emission factors of different pollutants can be calculated with respect to the average speed.

\subsection{Speed-density model}\label{sec: 3.5}

As mentioned in Section \ref{sec: 1}, a vehicle can deliver multiple passengers at each time in a ride-sharing system, enabling the system to realize comparable service rates with smaller fleet sizes. This, in turn, alleviates traffic congestion as well as increases traffic speed, due to the fewer number of on-road vehicles, which can further improve the transportation efficiency of the system. Hence, it is necessary to consider the ever-changing traffic speeds when quantifying the implications of ride-sharing services.

In this study, we adopt Greenshield's model, the simplest classic macroscopic traffic flow model proposed by Greenshield's in 1935, which assumes speed and density are linearly related under uninterrupted flow conditions, to calculate the traffic speeds of vehicles. The basic formula of Greesshield's model is as follows:

\begin{equation}
    \label{eq: Greenshield model}
    u = u_{0}(1-k/k_{j}),
\end{equation}
where $u$ denotes the traffic speed, $u_{0}$ the free-flow speed, $k$ the density, and $k_{j}$ the traffic jam density. The traffic jam density $k_{j}$ can be set as 100 vehicles/km/lane \citep{liu2012temporal}. According to the Amap' (one of the largest online map companies in China, affiliated with Alibaba) annual transportation report 2022\footnote{\url{https://report.amap.com/download.do} (in Chinese)}, the free-flow speed in Chengdu is around 45 km/h. The density $k$ consists of two components, basic density $k_b$ and ride-sourcing density $k_r$. The basic density $k_b$ refers to the density of on-road vehicles excluding ride-sourcing vehicles, such as buses and private cars. The ride-sourcing density $k_r$ refers to the density of the ride-sourcing vehicles that are simulated in this study. According to Chengdu transportation report 2021\footnote{\url{https://www.cdipd.org.cn/index.php?m=content&c=index&a=show&catid=63&id=385} (in Chinese)}, the ride-sourcing density $k_r$ approximately accounts for 10 - 15\% of the total density $k$. In this study, we assume the basic density $k_b$ remains constant because the real-time traffic flow data is inaccessible for now. This assumption is reasonable since this study focuses on the average values of traffic speeds and carbon emissions \citep{yan2020quantifying}.

This study adopts Dijkstra’s algorithm \citep{dijksta1959note} to estimate the shortest routes from origins to destinations of requests based on the road network. However, since this study accounts for traffic congestion when modeling vehicle behaviors, resulting in fluctuating vehicle speeds due to shifts in traffic density, the shortest routes are calculated using the on-road traffic time, rather than the road length. Specifically, this study adds an attribute, i.e., travel time, to each road, which is the weight of the road when calculating the shortest routes. Moreover, the system updates the traffic density of each road once a vehicle leaves or enters the road, based on which the travel time on the road can be updated according to the speed-density model.

\subsection{Simulation settings and measurement metrics}\label{sec: 3.6}

To simplify the analysis, this study focuses on two distinct scenarios: (1) a traditional ride-sourcing system without any ride-sharing options, where passengers do not share trips with others, and (2) a ride-sharing system where all passengers are willing to share trips. Any other scenarios in which a proportion of passengers are willing to share can be considered a combination of these two extremes. Moreover, unlike most existing literature that only allows at most two passengers to share a trip at each time, this study expands the analysis to include three different vehicle capacities, i.e., 2, 4, and 6. Therefore, there are 4 scenarios in total in this study: non-sharing ride-sourcing (denoted as NS), ride-sharing with the capacity of 2 passengers (denoted as RS2), ride-sharing with the capacity of 4 passengers (denoted as RS4), and ride-sharing with the capacity of 6 passengers (denoted as RS6). As mentioned earlier, ride-sharing services can achieve a similar service rate (SR) with a smaller fleet size, resulting in a reduction in traffic emissions and alleviating traffic congestion for a given level of demand. Therefore, this study calculates and compares traffic emissions and traffic congestion among four scenarios for a given level of demand and required SR. SRs mainly depend on demand-to-supply ratios, reflecting different ride-hailing markets, such as under-supply, balanced, and over-supply. For instance, when the demand-to-supply ratio is high, indicating under-supply, the SR should be relatively low. Consequently, this study further examines the performances of the four systems at different required SR levels, i.e., 50\%, 60\%, 70\%, 80\%, and 90\%.

This study uses system efficiency, traffic emissions, and traffic congestion as metrics to measure the overall performances of the four scenarios. The metrics are calculated as the average value of all observations and presented as relative values rather than absolute values. For example, the widely used metric, VMT, is not considered in this study as VMT is an absolute value that changes with fleet size. However, fleet size may vary significantly in different cities. Instead, the delivery efficiency factor (DEF) which refers to the ratio of VMT to passenger miles delivered (PMD) is used in this study to measure the required VMT for each PMD. Relative values are more typical and can be extrapolated to other scenarios.

First, this study adopts the SR and DEF to measure the system efficiency under various scenarios. The SR can be calculated as follows:

\begin{equation}
    \label{eq: SR}
    SR = \frac{n}{N},
\end{equation}
where $n$ and $N$ denote the number of served passengers and the total number of passengers, respectively. As introduced above, DEF can be calculated as:

\begin{equation}
    \label{eq: DEF}
    DEF = \frac{VMT}{PMD}.
\end{equation}
In particular, SR can be used to measure the overall performance of the system, while DEF can be used to measure the efficiency of delivering passengers. 

For traffic emissions, the amount of emissions may vary significantly between different scenarios due to the different settings such as fleet size. As a result, similar to DEF, this study uses the passenger emission factor (PEF) to measure the emissions of various pollutants. PEF of Pollutant $k$ can be calculated as follows:

\begin{equation}
    \label{eq: PEF}
    PEF_k = \frac{E_k}{PMD}.
\end{equation}
Remind that $E_k$ denotes the amount of pollutant $k$ emitted by all vehicles. This metric can represent the emission cost for delivering each passenger mile, which is easier to understand and more convenient for comparison between various scenarios.

Third, this study adopts the delay factor (DF) to measure the traffic congestion under different scenarios \citep{erhardt2019transportation}. The delay factor refers to the ratio of congested travel time to free-flow travel time. Hence, the higher the value of the delay factor, the longer the travel delay, i.e., the more congested the road is. At an observation time point $t$, for each road $j$, the delay factor $DF_{t,j}$ can be calculated as:

\begin{equation}
    \label{eq: delay factor}
    DF_{t, j} = \frac{l_j/u_{t,j}}{l_j/u_0} = \frac{u_0}{u_{t,j}},
\end{equation}
where $u_{t,j}$ denotes the average space speed of road $j$ at time point $t$. Remind that $l_j$ and $u_0$ are the length of the road $j$ and the free-flow speed, respectively. Therefore, the average delay factor among all roads for all observations $D$ can be calculated as follows:

\begin{equation}
    \label{eq: delay}
    DF = \frac{1}{T}\frac{1}{J}\sum_{t}{\sum_{j}{D_{t, j}}},
\end{equation}
where $T$ and $J$ denote the number of observation time points and roads, respectively.

To accelerate the simulation process, this study randomly downsamples approximately 30\% of the requests, which does not affect the spatial and temporal distributions of requests. The simulation time step as well as the batch-matching time interval are set to 2 seconds. At the end of each batch-matching time frame, this study calculates the DF of all roads, and when a vehicle arrives at the end of a road, this study calculates the traffic emissions of this vehicle on the road. In addition, several assumptions are made for the simulation. First, to simplify the model, car-following behaviors are not considered, and the speeds of vehicles on the road change instantaneously when the density of the road changes. This assumption is reasonable since this study only considers the average speeds of on-road vehicles when calculating the emission factor and delay factor. Second, idle vehicles are assumed to be cruising randomly within their surrounding areas. Once the system assigns a passenger to a driver, the driver stops cruising and drives to pick up and deliver the passenger. It should be noted that the VMT for cruising is included in the total VMT since idle vehicle cruising is essential for ride-sourcing or ride-sharing systems to serve more passengers, which is the fundamental objective of a mobility system. Third, this study assumes that each request includes only one passenger; thus there is no difference between a passenger and a passenger request in this study.


\section{Results and discussions}\label{sec: 4}
\subsection{Overall emission reductions and traffic congestion alleviation from ride-sharing}\label{sec: 4.1}

The experimental results are presented in Fig. \ref{fig: result1}, with the required SRs ranging approximately from 40\% to 95\%. Additionally, the system efficiency improvement, traffic congestion alleviation, and traffic emission reductions resulting from ride-sharing at different levels of SR are calculated via interpolation and shown in Fig. \ref{fig: result2}. In general, both ride-sourcing and ride-sharing systems need to pay a higher cost to achieve a higher SR. This is mainly because, when the system is under-supply, it is easier for idle vehicles to find passengers, resulting in fewer cruising distances, higher system efficiencies, fewer traffic emissions, and less traffic congestion. However, when the required SR increases, the corresponding required fleet size also increases, leading to more idle vehicles cruising longer distances to find passengers, which results in more traffic emissions and heavier traffic congestion.

\begin{figure}[!htbp]
    \centering
    \subfigure[]{\includegraphics[width=0.35\linewidth, height=0.35\linewidth]{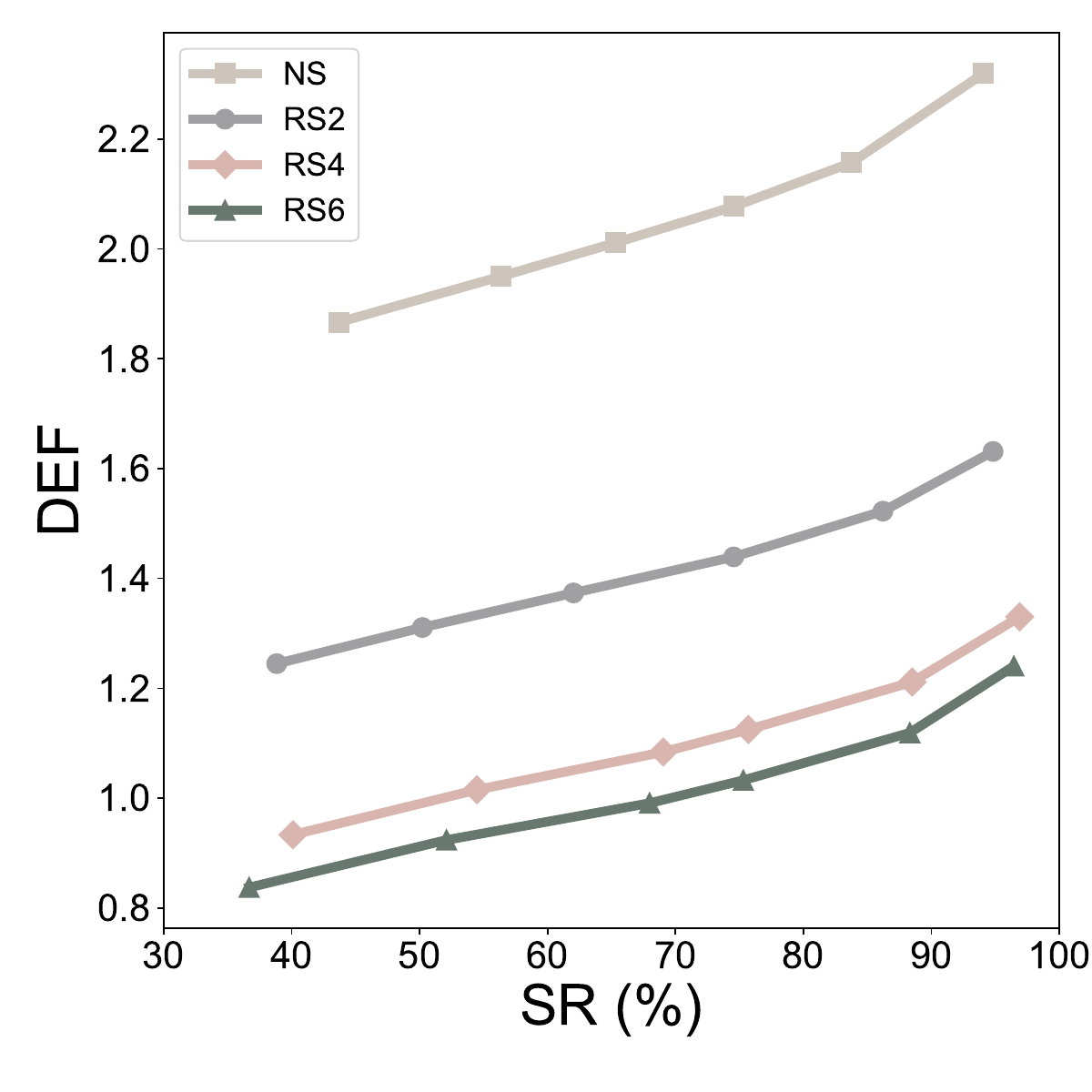}}
    \subfigure[]{\includegraphics[width=0.35\linewidth,height=0.35\linewidth]{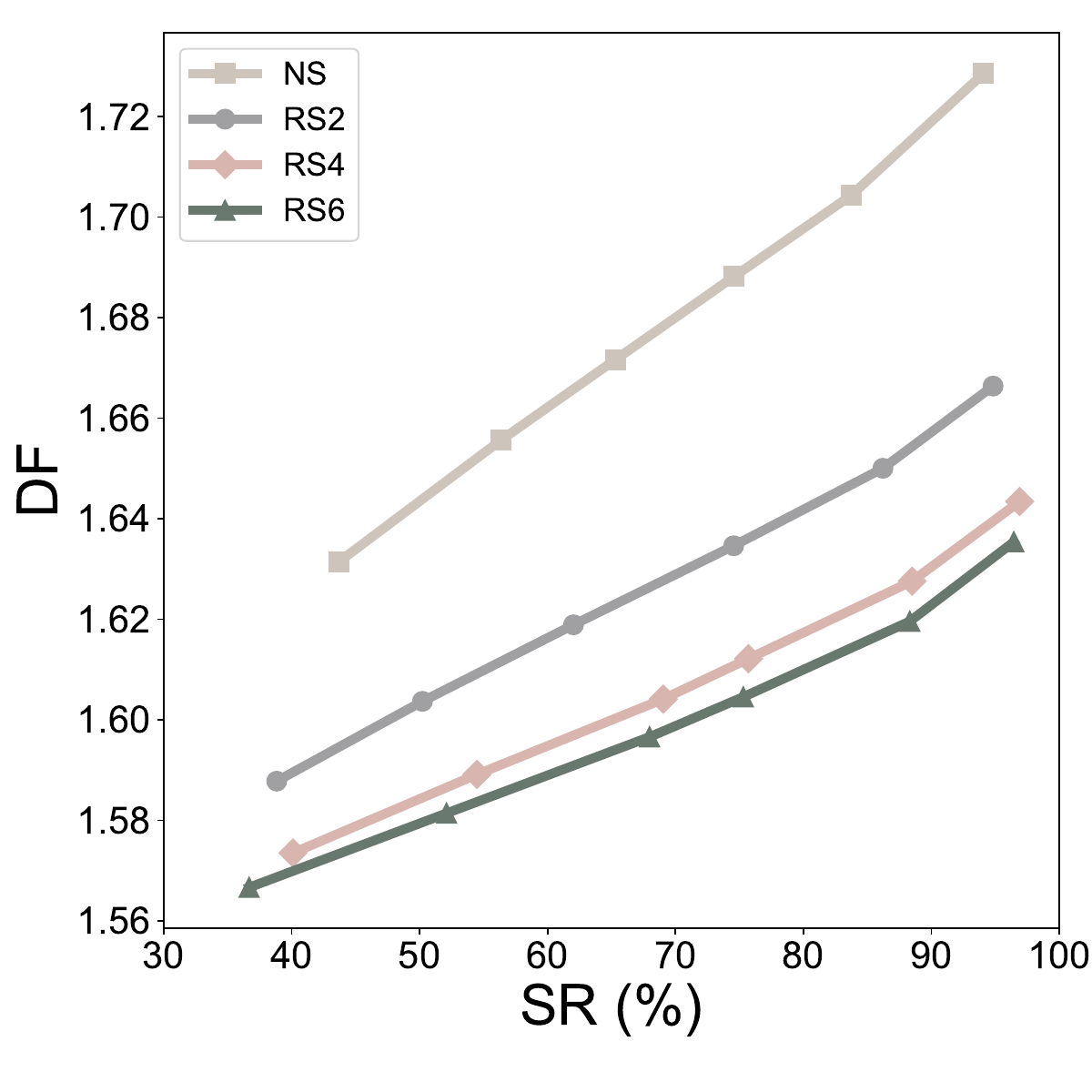}}
    \subfigure[]{\includegraphics[width=0.35\linewidth,height=0.35\linewidth]{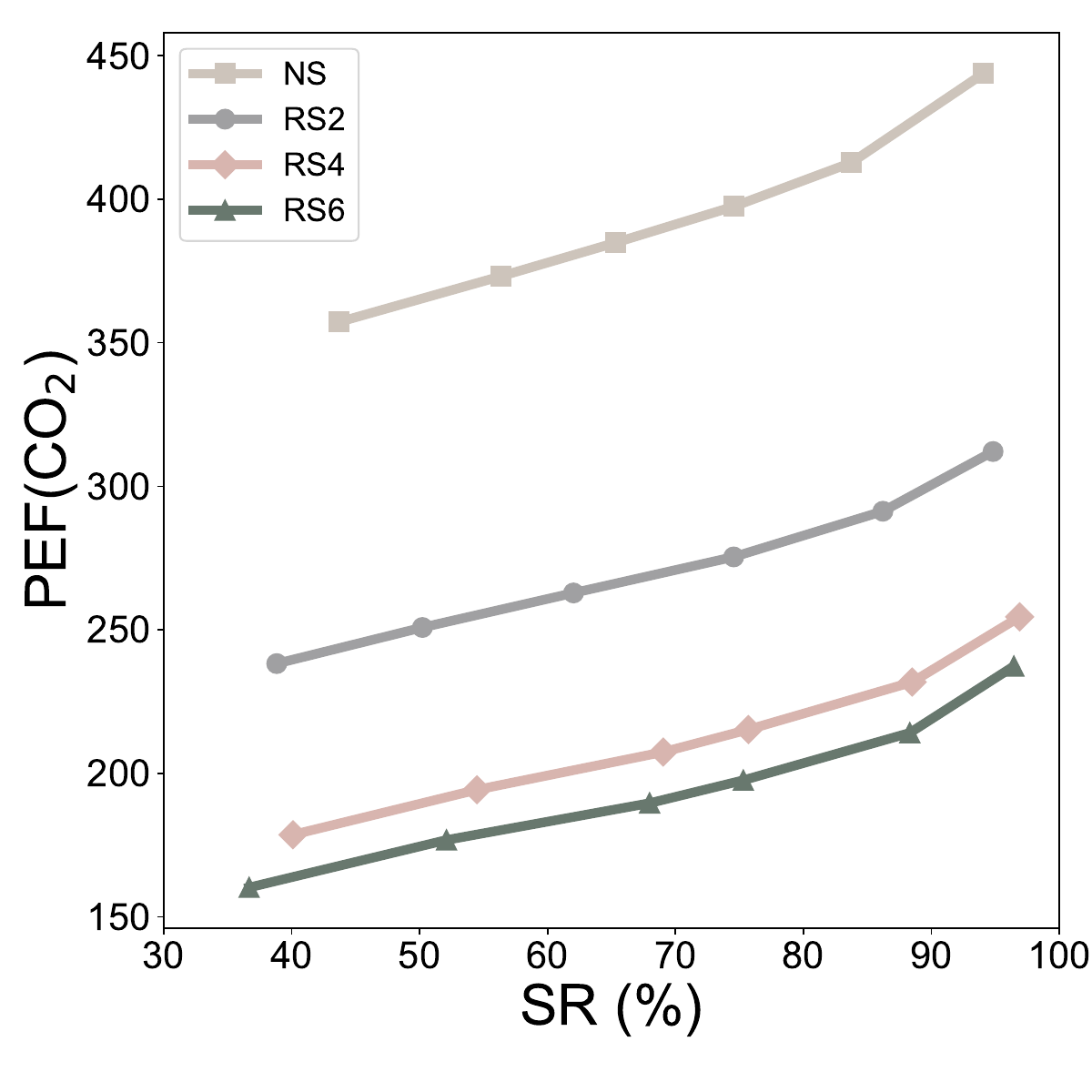}}
    \subfigure[]{\includegraphics[width=0.35\linewidth,height=0.35\linewidth]{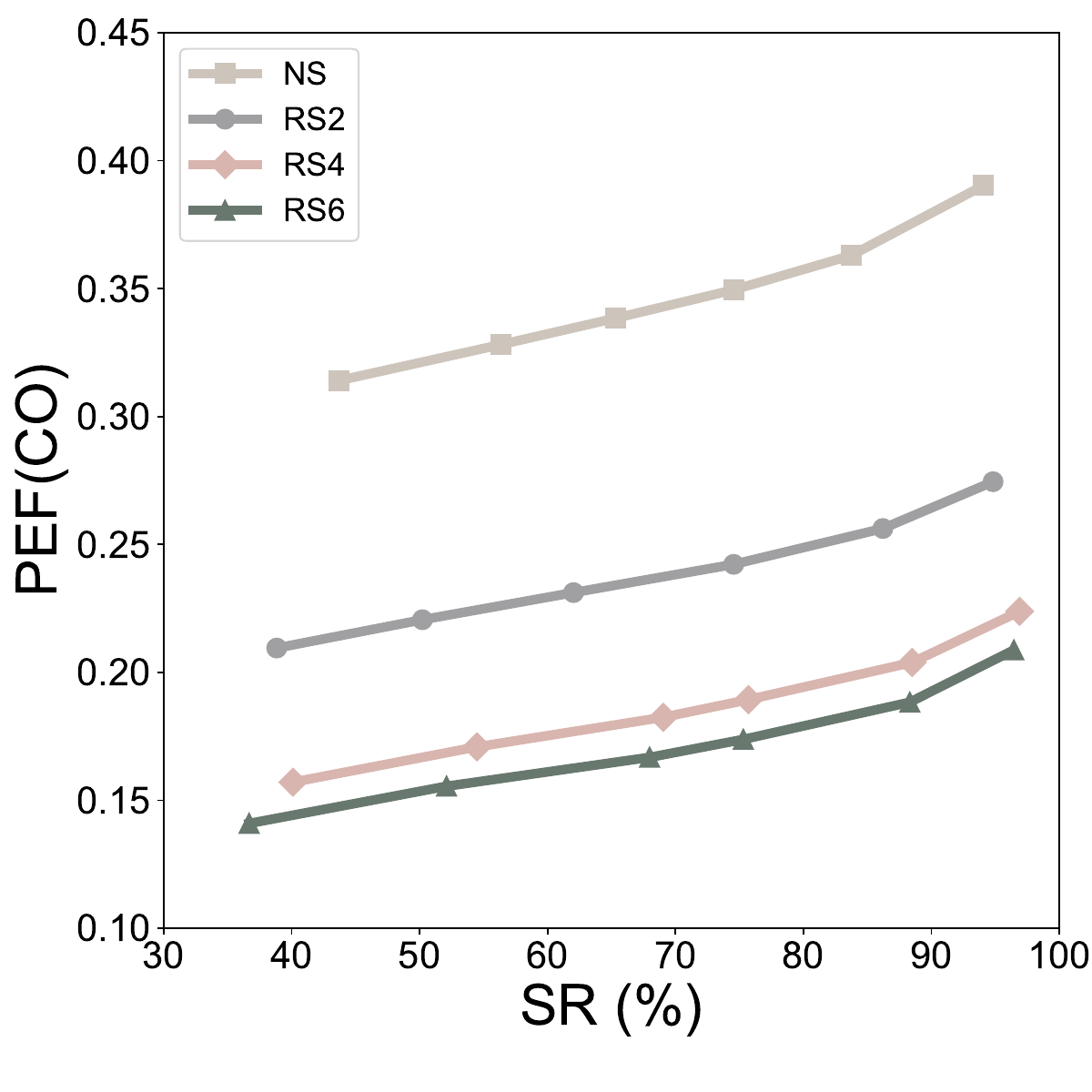}}
    \subfigure[]{\includegraphics[width=0.35\linewidth,height=0.35\linewidth]{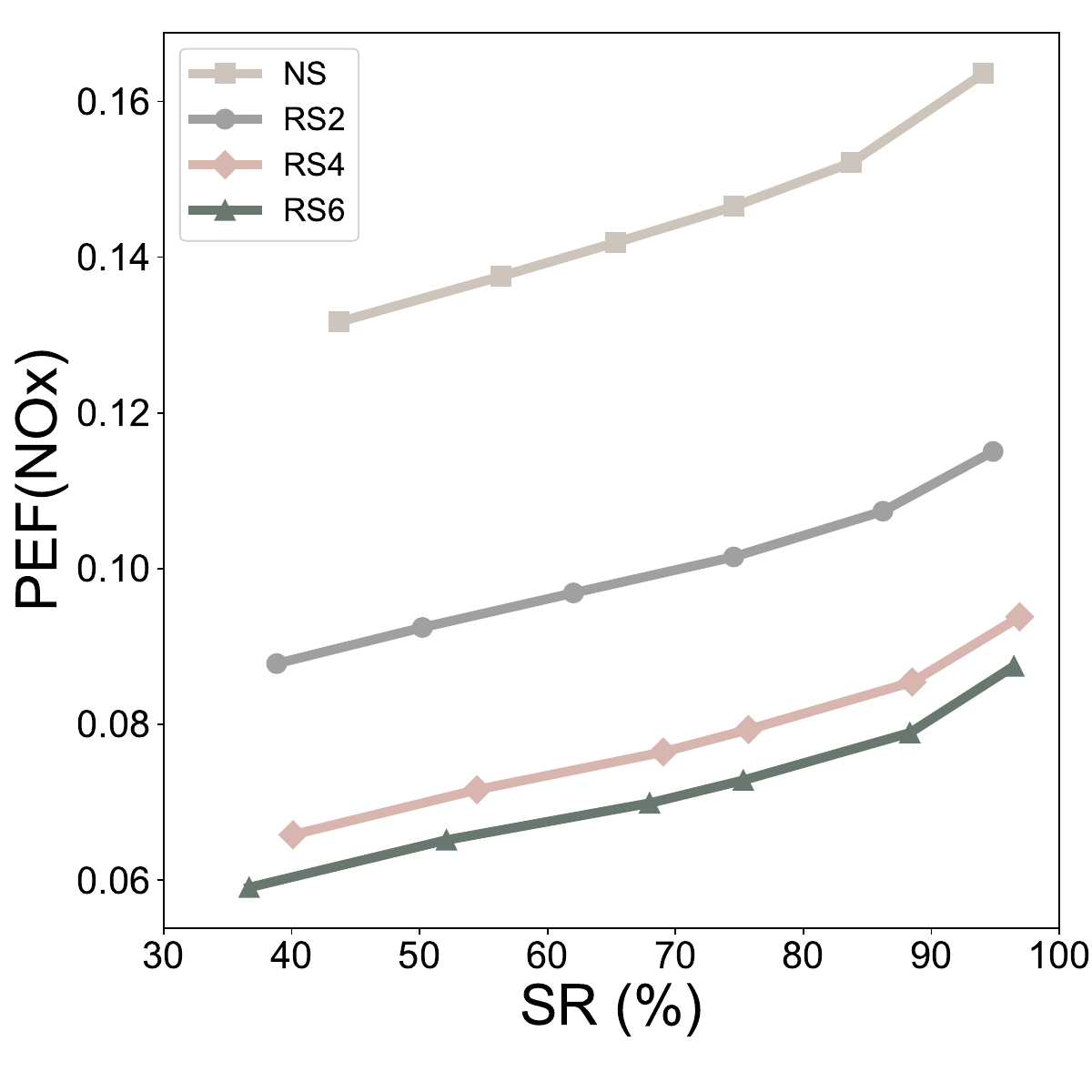}}
    \subfigure[]{\includegraphics[width=0.35\linewidth,height=0.35\linewidth]{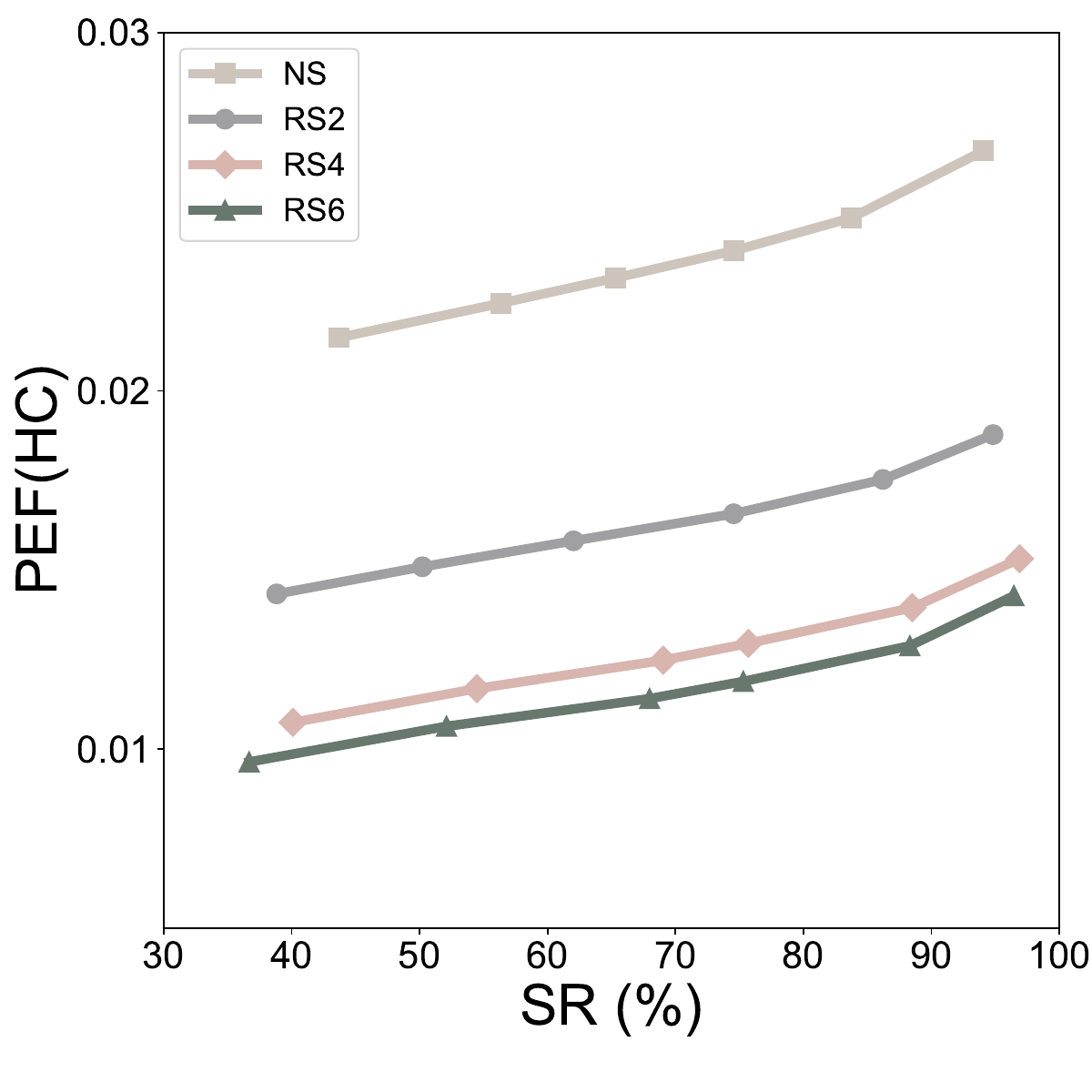}}
    \caption{Experimental results on the system efficiency, traffic congestion, and traffic emissions at different levels of SRs.}
    \label{fig: result1}
\end{figure}

\begin{figure}[!htbp]
    \centering
    \subfigure[]{\includegraphics[width=0.35\linewidth, height=0.35\linewidth]{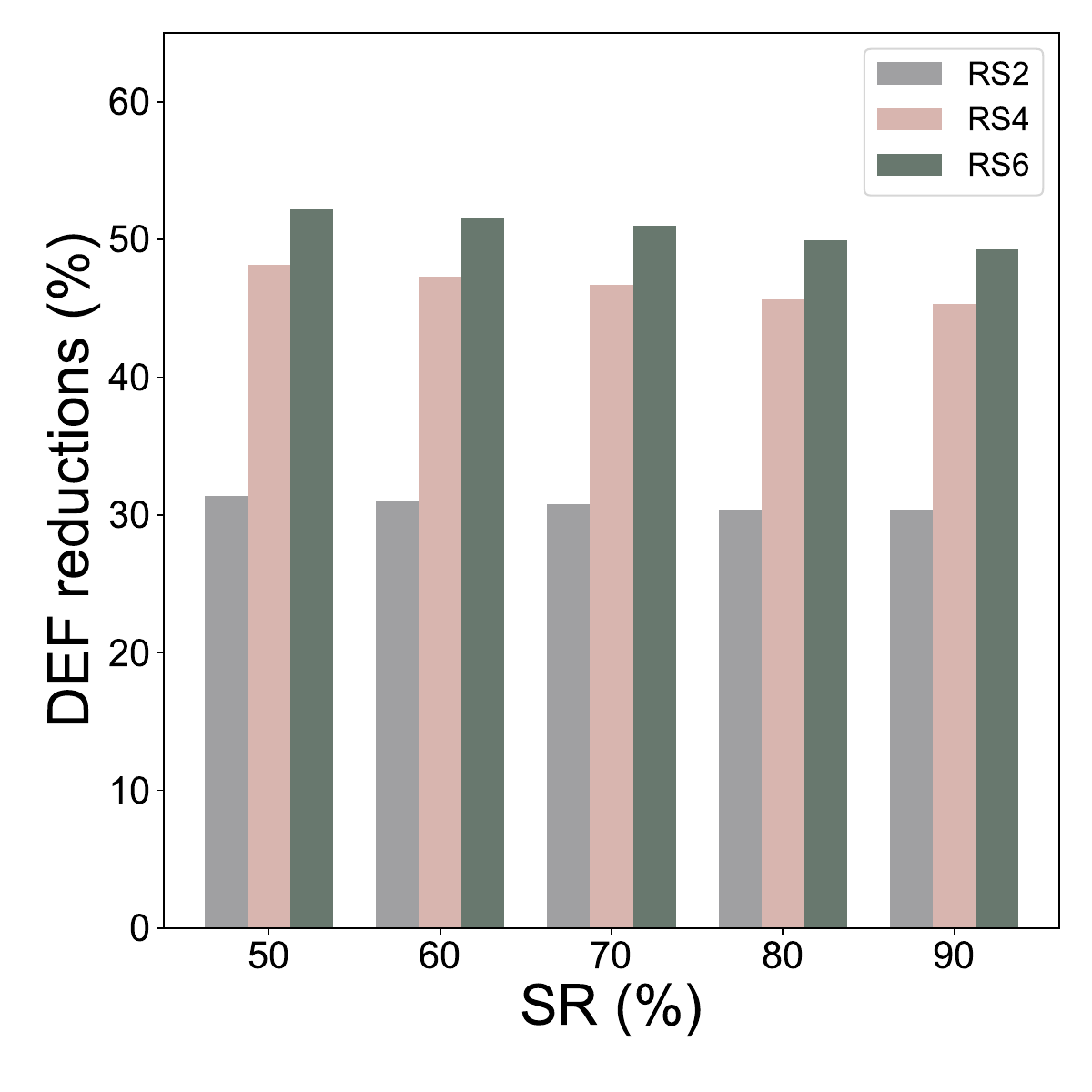}}
    \subfigure[]{\includegraphics[width=0.35\linewidth,height=0.35\linewidth]{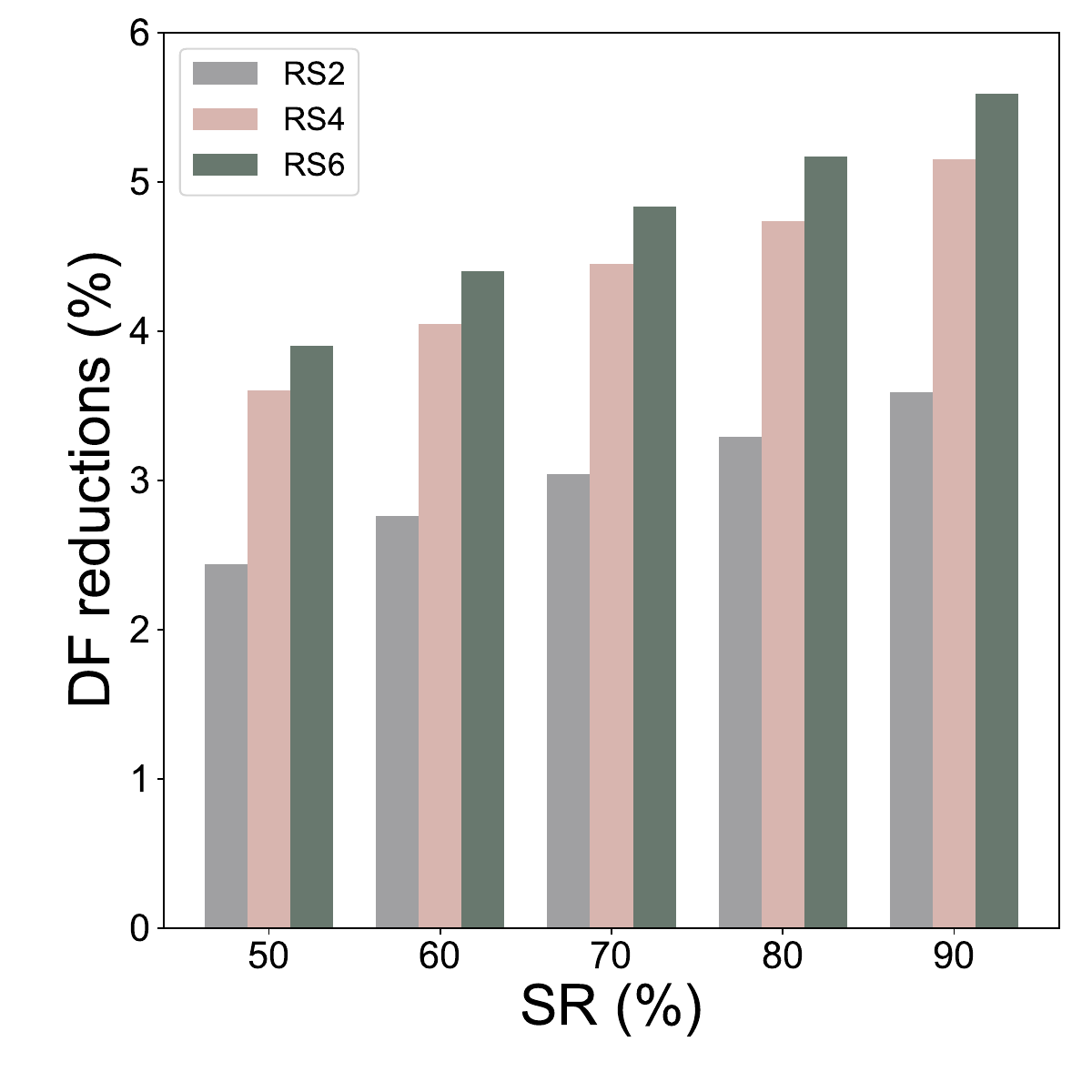}}
    \subfigure[]{\includegraphics[width=0.35\linewidth,height=0.35\linewidth]{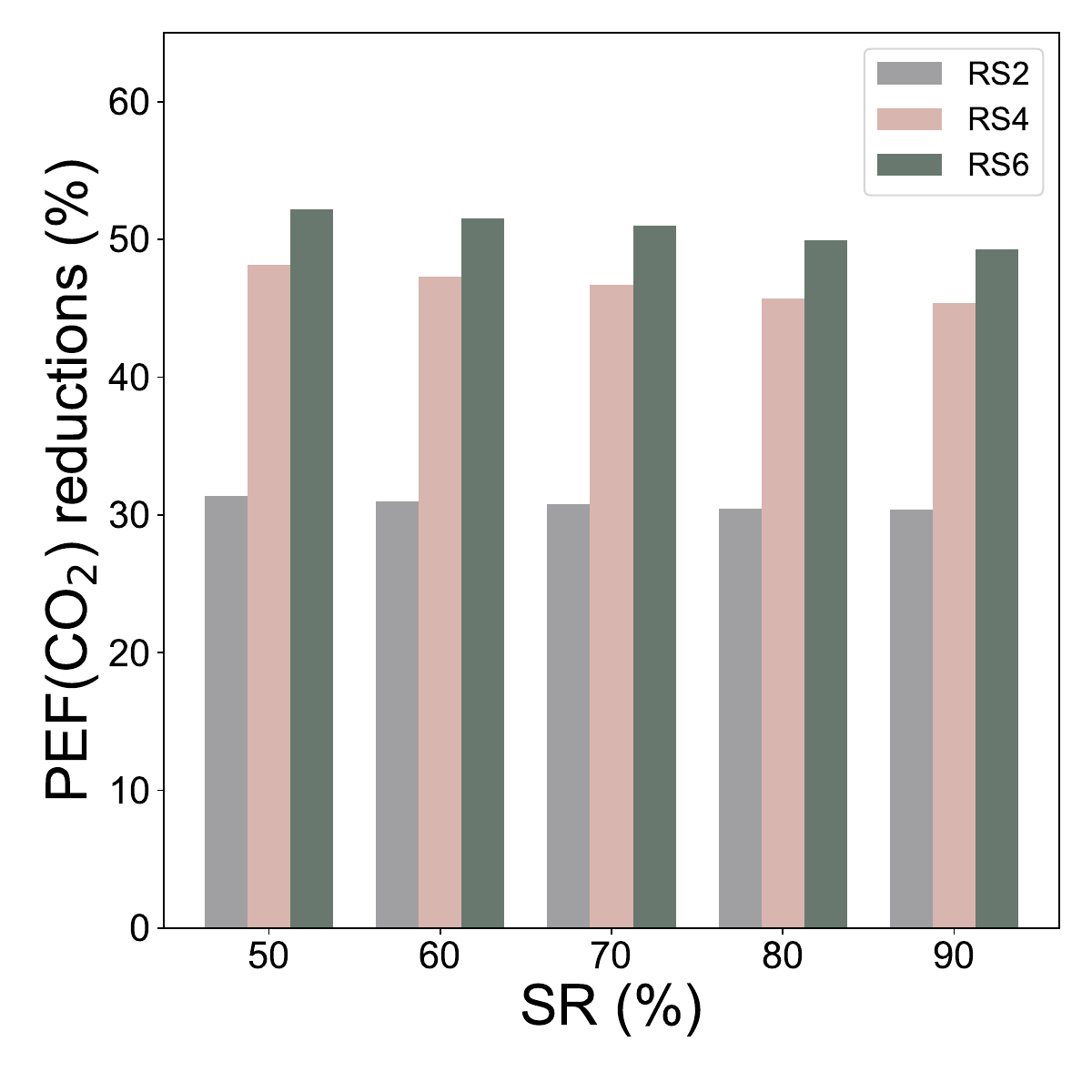}}
    \subfigure[]{\includegraphics[width=0.35\linewidth,height=0.35\linewidth]{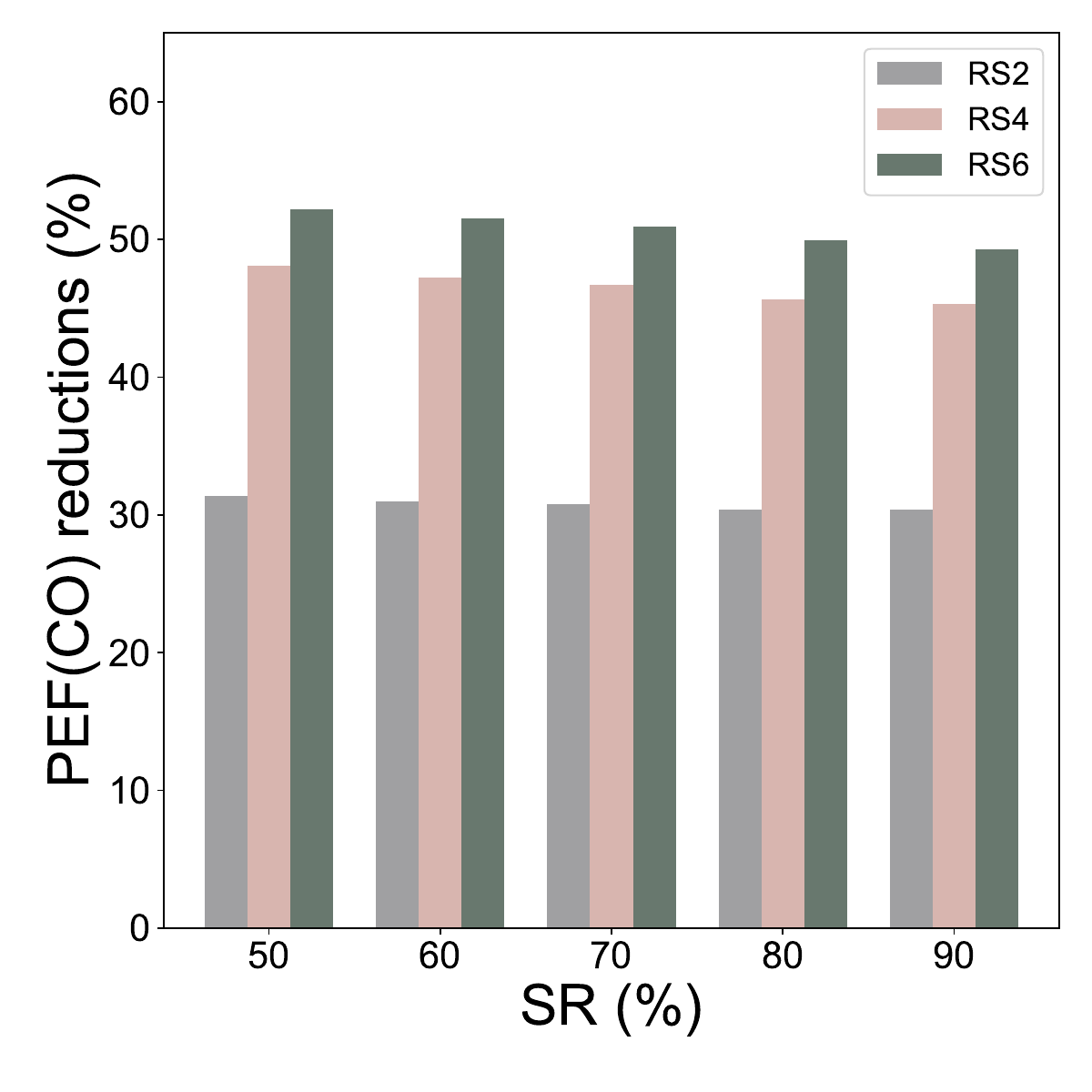}}
    \subfigure[]{\includegraphics[width=0.35\linewidth,height=0.35\linewidth]{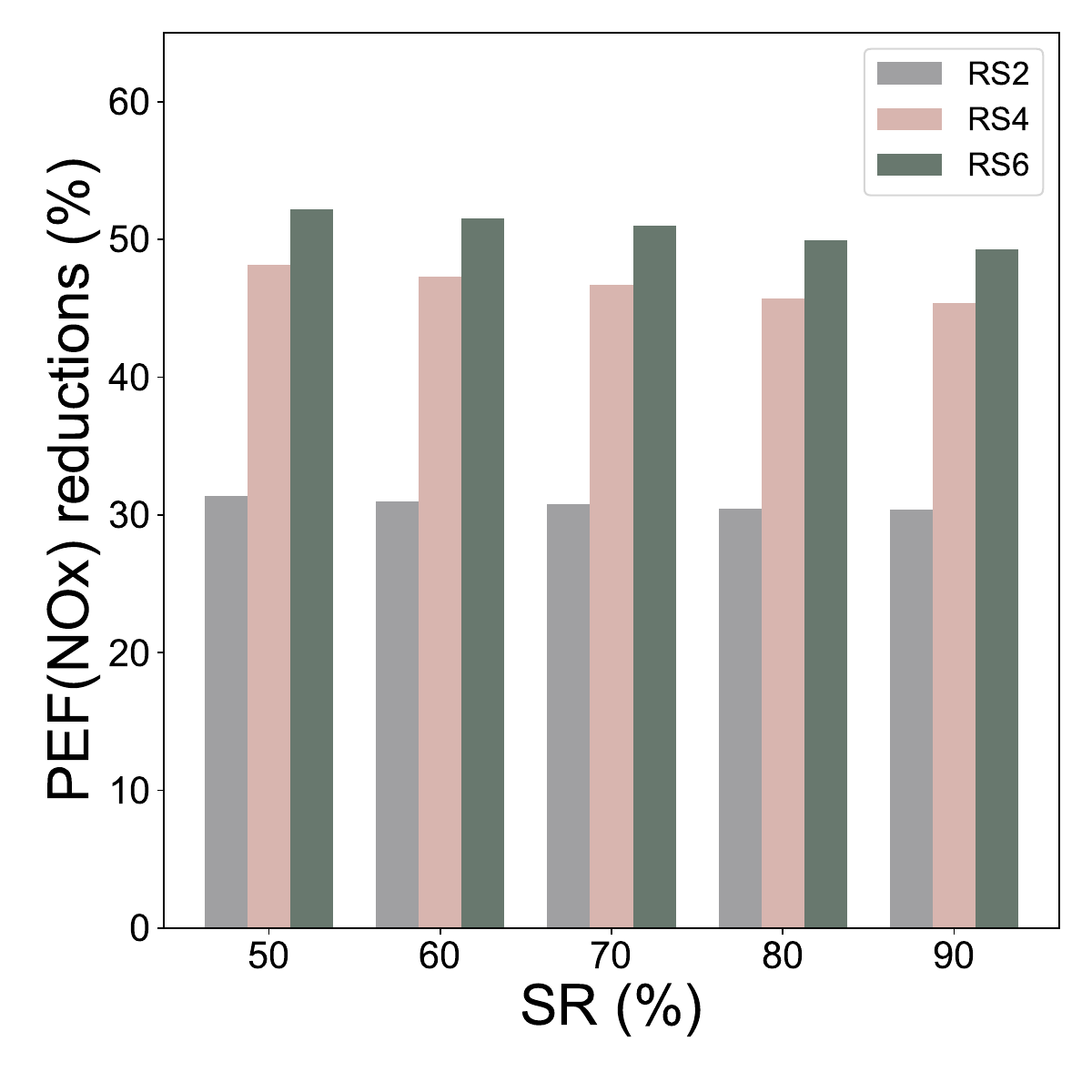}}
    \subfigure[]{\includegraphics[width=0.35\linewidth,height=0.35\linewidth]{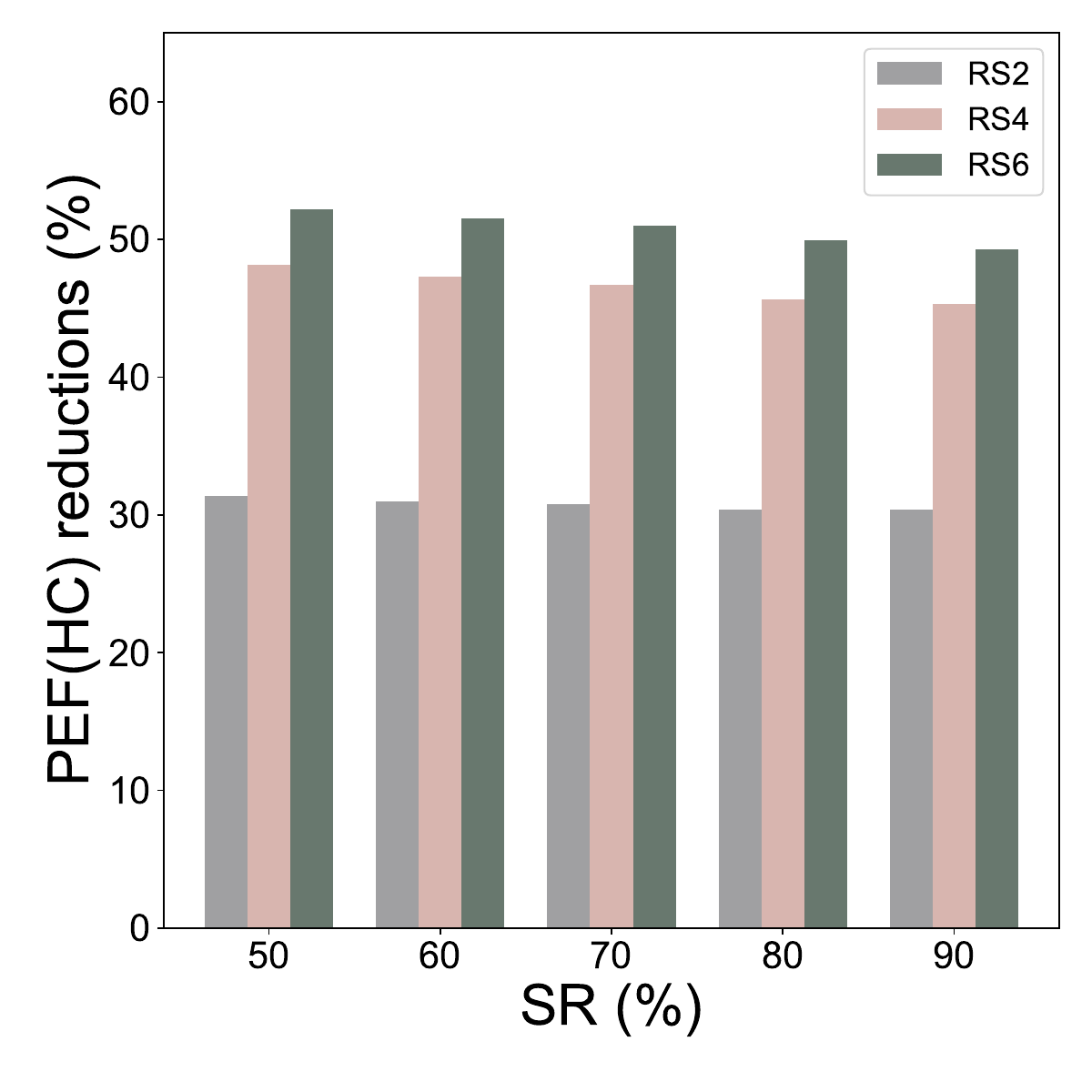}}
    \caption{System efficiency improvement, traffic congestion alleviation, and traffic emission reductions resulting from ride-sharing at different levels of SRs.}
    \label{fig: result2}
\end{figure}

In terms of traffic congestion, at different levels of required SRs, ride-sharing scenarios SR2, SR4, and SR6 can reduce the values of DF by 2.4 - 3.6\%, 3.6 - 5.2\%, and 3.9 - 5.6\%, respectively (Fig. \ref{fig: result2} (b)). However, it should be noted that ride-sourcing vehicles only account for 10 - 15\% of the total traffic flow, leading to a relatively small improvement in traffic congestion resulting from ride-sharing. In addition, when the required SR increases, the reductions in traffic congestion resulting from ride-sharing are more significant (Fig. \ref{fig: result2} (b)). For instance, when the SR is at 50\%, the reductions of DF resulting from the three ride-sharing scenarios are 2.4\%, 3.6\%, and 3.9\%, respectively. However, when the required SR increases to 90\%, the reductions of DF increase to 3.6\%, 5.2\%, and 5.6\%, respectively. This is because, for a given level of demand, when the required SR increases, the corresponding fleet size also increases. Therefore ride-sharing services can reduce a larger absolute number of vehicles, resulting in a more significant decrease in traffic flow, since this study assumes the basic flow is instant.

As for traffic emissions, including CO$_2$, CO, NOx, and HC, ride-sharing scenarios SR2, SR4, and SR6 can reduce the values of PEF by approximately 30\%, 45\%, and 50\%, respectively (Fig. \ref{fig: result2} (c-f)). In other words, compared to a non-sharing ride-sourcing system, the three ride-sharing systems can achieve a PMD with approximately 70\%, 55\%, and 50\% of traffic emissions. These reductions are mainly brought by the improvement of traffic efficiency resulting from ride-sharing, as shown in Fig. \ref{fig: result2} (a). Ride-sharing can reduce the required VMT to achieve a PMD, thereby reducing traffic emissions. Furthermore, when the required SR increases, the reductions in traffic emissions provided by ride-sharing decrease slightly (Figs. \ref{fig: result2} (c) - (f)). This is because when the required SR is low, indicating that the system is under-supply, it is easier for passengers to share trips, leading to more significant reductions in traffic emissions. However, as the required SR increases, the corresponding required fleet size increases, then it becomes more difficult to match passengers with compatible travel routes and schedules, leading to a diminishing impact on traffic emissions.

\subsection{Emission reductions from traffic speed increase effect}\label{sec: 4.2}

Ride-sharing services can reduce the required fleet size for a given level of demand, resulting in decreased traffic density and increased traffic speeds compared to non-sharing ride-sourcing services. The speed increase resulting from ride-sharing can further impact traffic emissions since the emissions of various pollutants are speed-depended. This study calculates the additional emission reductions from the traffic speed increase effect provided by ride-sharing services, as illustrated in Fig. \ref{fig: SpeedEffect}. In addition to CO, emissions of CO$_2$, NOx, an HC can be further reduced, with reductions ranging from approximately 0.38\% - 0.85\%, 0.16\% - 0.38\%, and 0.02 - 0.047\%, respectively. These findings are consistent with those reported by \cite{yan2020quantifying}. These percentage reductions are relatively small because ride-sourcing vehicles constitute only a small proportion of the total traffic flow. Therefore, the increase in traffic speed due to reduced fleet size in ride-sharing systems may not be noticeable.

\begin{figure}[!h]
    \centering
    \subfigure[CO$_2$]{\includegraphics[width=0.4\linewidth,height=0.4\linewidth]{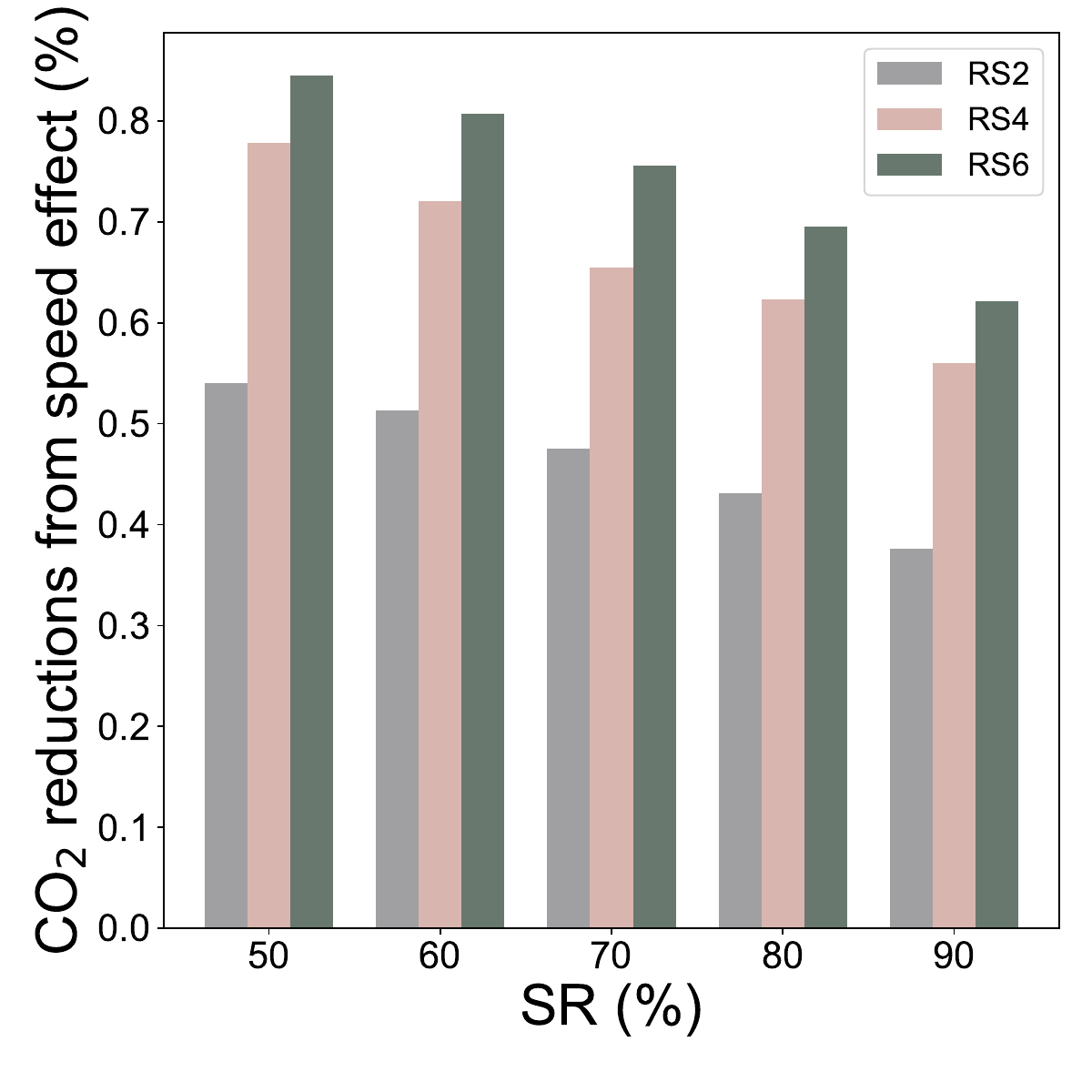}}
    \subfigure[CO]{\includegraphics[width=0.4\linewidth,height=0.4\linewidth]{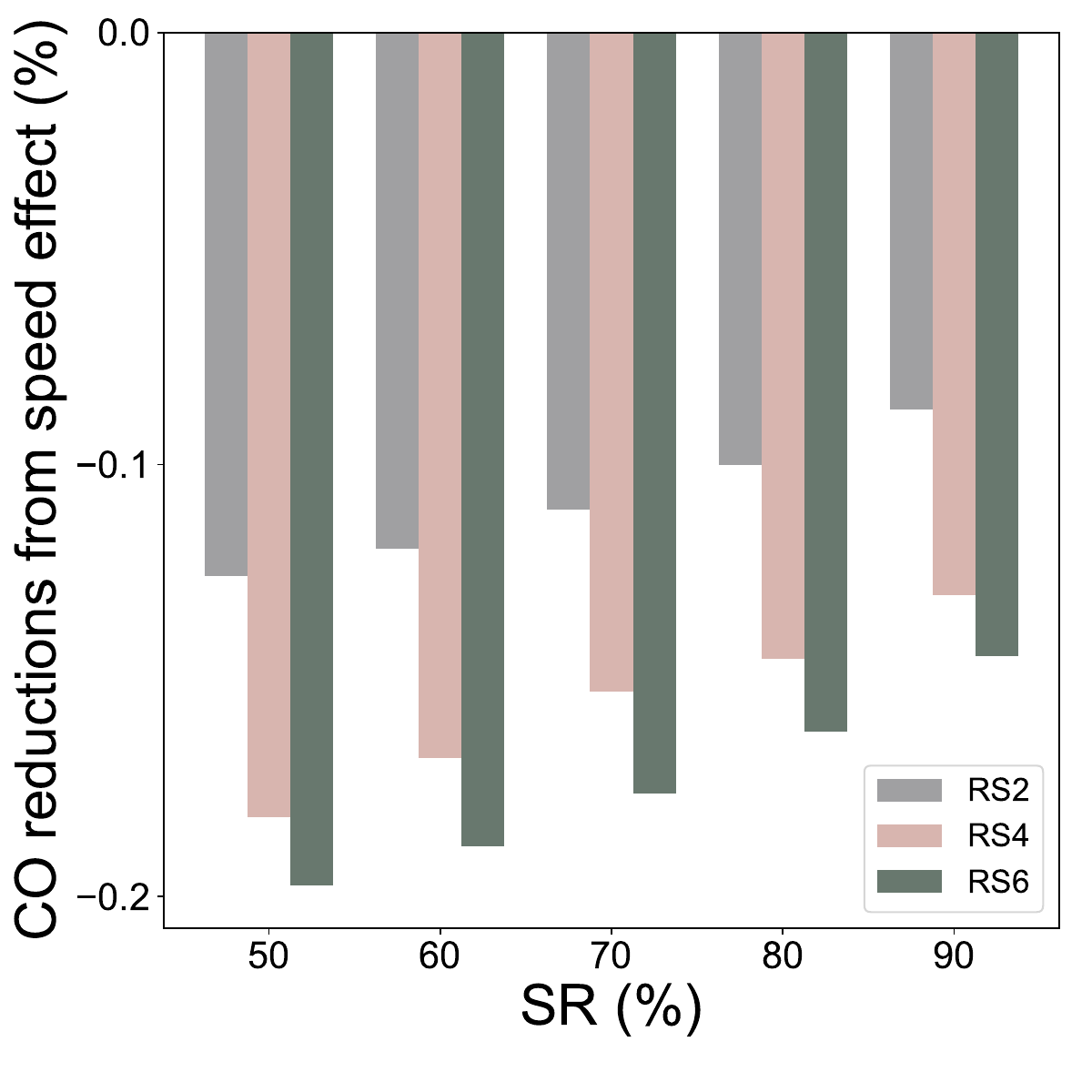}}
    \subfigure[NOx]{\includegraphics[width=0.4\linewidth,height=0.4\linewidth]{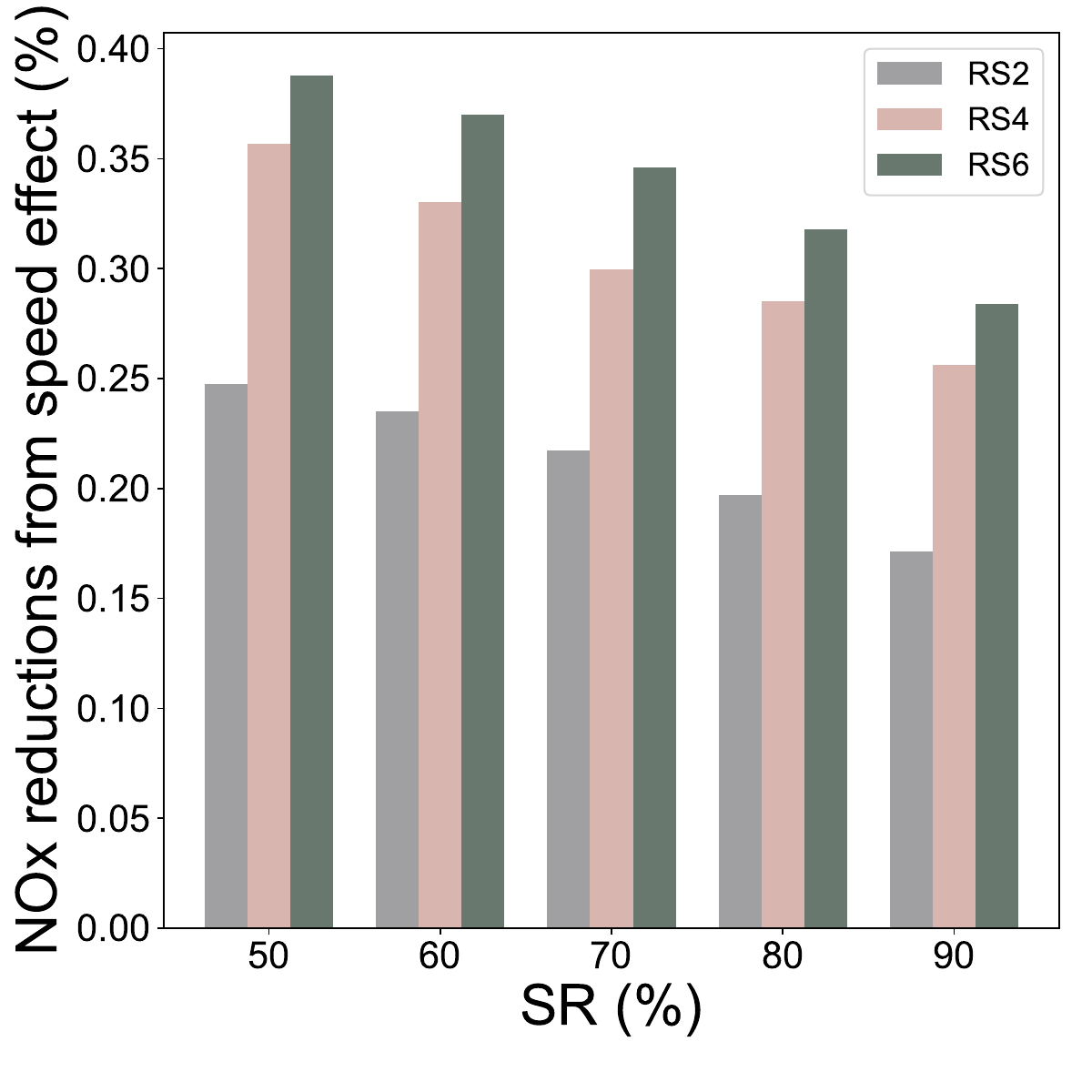}}
    \subfigure[HC]{\includegraphics[width=0.4\linewidth,height=0.4\linewidth]{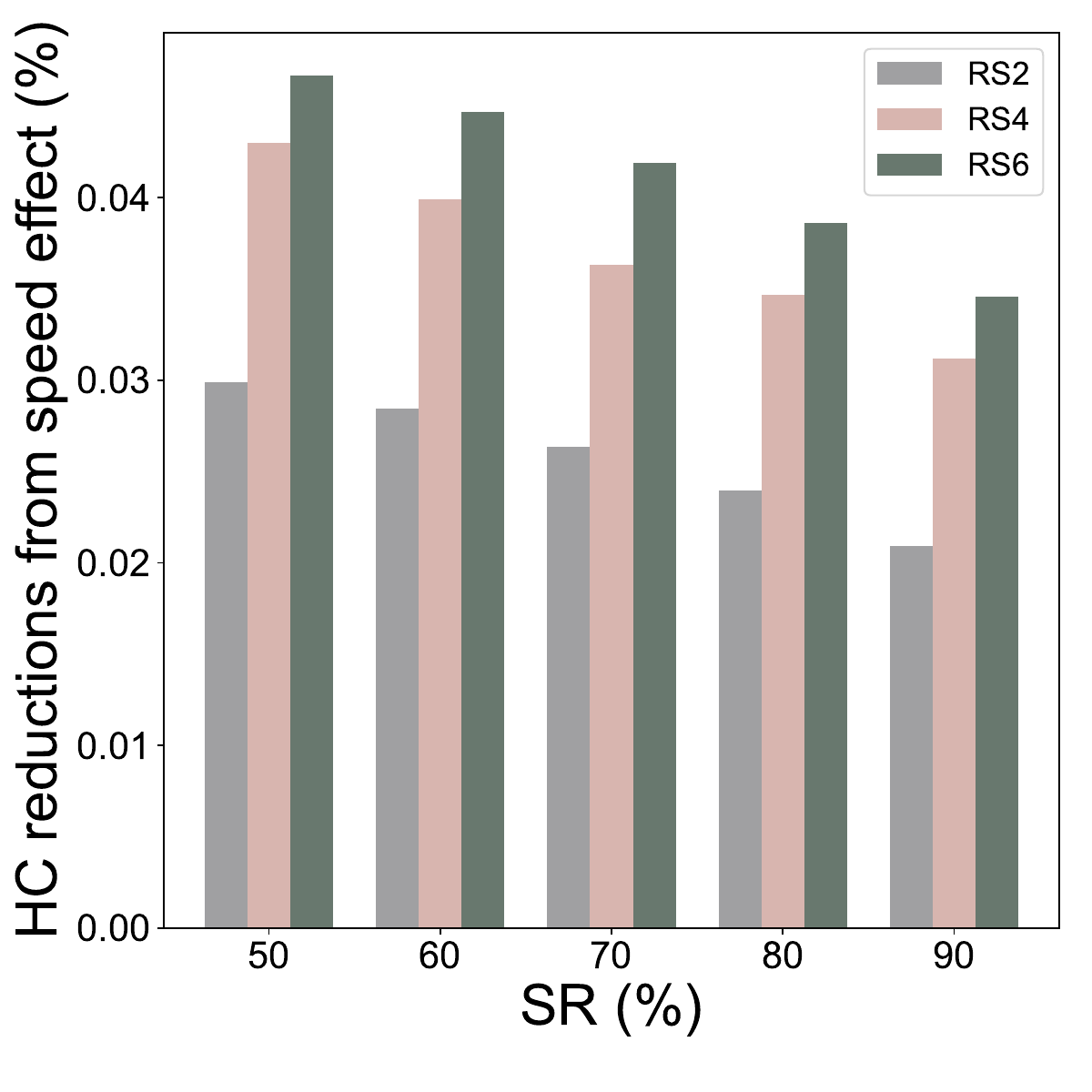}}
    \caption{Traffic emission reductions from traffic speed increase effect.}
    \label{fig: SpeedEffect}
\end{figure}

While the further improvements in emission reductions achieved by traffic speed increase may not be significant in this study, the results provide a valuable reference for future research on reducing traffic emissions. Specifically, the findings suggest that traffic emissions can be reduced not only by decreasing the required fleet size to meet a given level of demand but also by increasing traffic speeds. Furthermore, in future mobility systems where all vehicles are autonomous and connected, ride-sharing programs can potentially be applied to all vehicles, resulting in a significant decrease in traffic flow and a corresponding increase in traffic speed. Therefore, it is possible to achieve a substantial further reduction in traffic emissions through the implementation of such ride-sharing programs.

\subsection{Spatial pattern analysis}\label{sec: 4.3}

The study area contains an obvious hot zone with high demand, located in the middle of the area with a longitude from 104.4 to 104.12 and a latitude from 30.64 to 30.70. To analyze the spatial patterns of the impact of ride-sharing services, this study calculates the traffic emission reductions and traffic congestion alleviation resulting from ride-sharing services in the hot zone and other areas, as shown in Table \ref{table3}. For the sake of conciseness, this section only analyzes the spatial patterns at 80\% of the SR level, with similar results at other SR levels displayed in the appendix. The results show that the impacts of ride-sharing services on reducing traffic emissions and alleviating traffic congestion in the hot zone are more significant than those in other areas. For example, in the hot zone, ride-sharing scenarios RS2, RS4, and RS6 can reduce CO$_2$ emissions by 32.6\%, 49.7\%, and 54.3\%, respectively. However, these reduction percentages decrease to 30.0\%, 45.5\%, and 50.3\%, respectively, in other areas. Similarly, the decreased DFs in three ride-sharing scenarios are 3.1\%, 4.5\%, and 4.9\%, respectively, in other areas. In contrast, the alleviation of traffic congestion provided by RS2, RS4, and RS6 is more significant in the hot zone, with reductions in DFs of 3.5\%, 5.0\%, and 5.5\%, respectively. This can be explained by the spatial aggregation of trip origins, as demonstrated in Fig. \ref{fig: SpaDis} (a). Most ride requests are concentrated in the hot zone, increasing the probability of passengers finding others with similar destinations to share a vehicle. This outcome results in a more significant decrease in the required fleet size to satisfy mobility needs, leading to more substantial reductions in traffic emissions and the alleviation of traffic congestion.

\begingroup
\setlength{\tabcolsep}{6pt} 
\renewcommand{\arraystretch}{1} 
\begin{table}[!htbp]
\caption{Comparison of the impacts of ride-sharing services on traffic emission reductions and traffic congestion alleviation in the hot zone and other areas at 80\% of SR.}
\begin{center}
\begin{tabular}{ c|c|c|c|c|c|c } 
 \hline
 & RS2 (other) & RS2(hot) & RS4 (other) & RS4(hot) & RS6 (other) & RS6(hot)\\
 \hline
CO$_2$ (\%) & 30.0 & 32.6 & 45.5 & 49.7 & 50.3 & 54.3 \\
CO (\%) & 30.2 & 32.7 & 45.4 & 49.8 & 50.3 & 54.3 \\
NOx (\%) & 30.0 & 32.7 & 45.5 & 49.8 & 50.6 & 54.4 \\
HC (\%) & 29.9 & 32.6 & 45.4 & 49.8 & 50.4 & 54.2 \\
DF (\%) & 3.1 & 3.5 & 4.5 & 5.0 & 4.9 & 5.5 \\

\hline
\end{tabular}
\label{table3}
\end{center}
\end{table}

To further analyze the spatial patterns of ride-sharing services' implications, this study aggregates the results across all observations and calculates the daily emissions of non-sharing ride-sourcing and the emission reductions from ride-sharing on each road. The emissions of non-sharing ride-sourcing and the emission reductions from ride-sharing on each road are then plotted as a scatter plot of ($x$,$y$), as shown in Fig. \ref{fig: reg80}. A linear regression without a constant is used to model the relationship between $x$ and $y$, and the regression results are also shown in Fig. \ref{fig: reg80}. The regressed equations indicate a strong linear relationship between $x$ and $y$, with an R$^2$ value of more than 0.9. This suggests that the traffic emission reductions from ride-sharing services are spatially correlated with emissions from non-sharing ride-sourcing services. In other words, ride-sharing can reduce more traffic emissions on roads with higher traffic emissions from non-sharing ride-sourcing, as shown in Fig. \ref{fig: vis80}. Moreover, the regression results demonstrate that ride-sharing systems SR2, SR4, and SR6 can reduce emissions of CO$_2$, CO, NOx, and HC by approximately 37.6\%, 55.2\%, and 59.3\%, 38.1\%, 56.4\%, and 61.6\%, 36.7\%, 54.2\%, and 58.9, and 36.9\%, 54.3\%, and 58.5\%, respectively.

\begin{figure}[!h]
    \centering
    \includegraphics[width=0.92\linewidth]{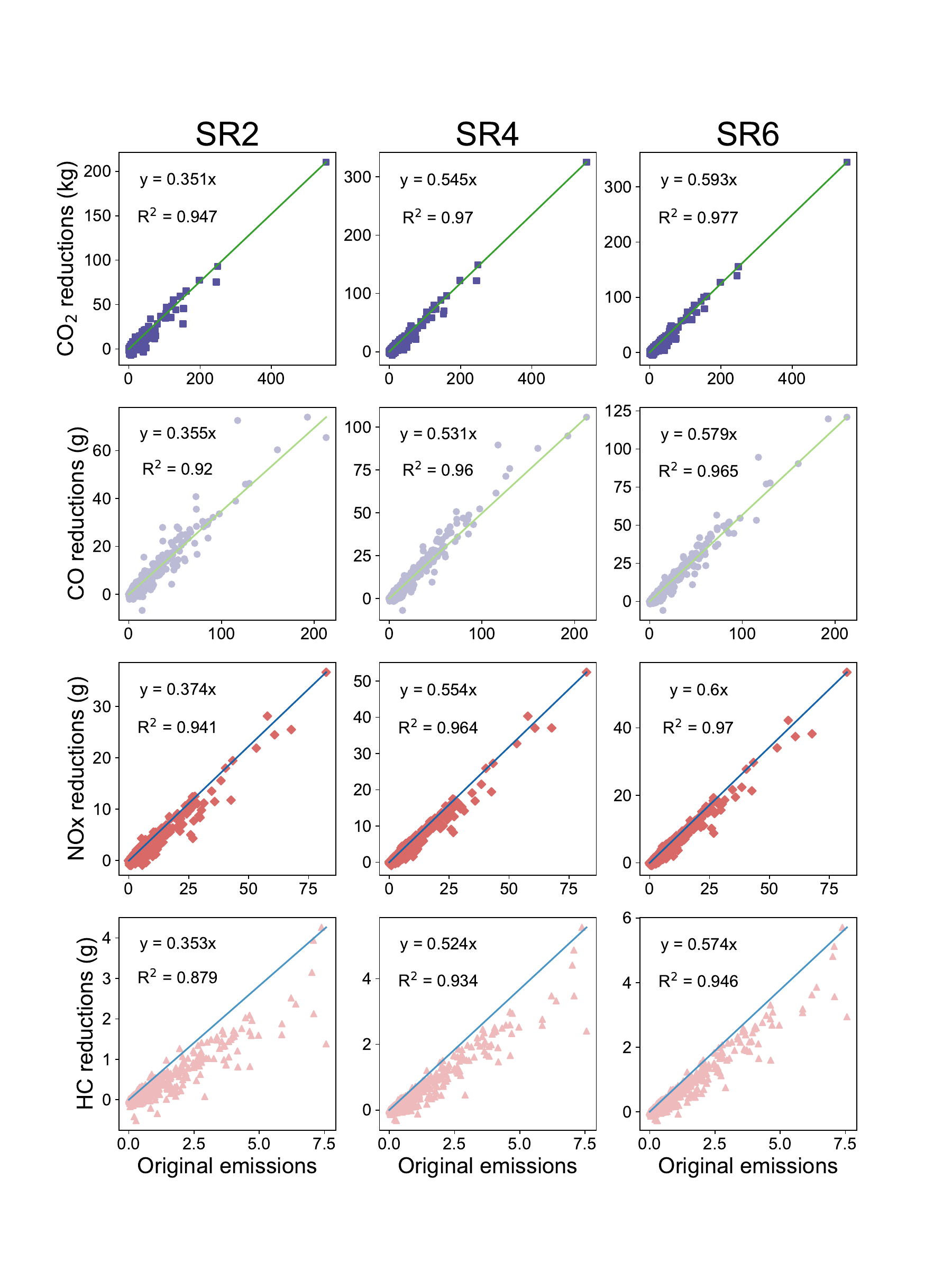}
    \caption{Relationship between reduced traffic emissions brought by ride-sharing and original emissions from non-sharing ride-sourcing on each road at 80\% of SR level.}
    \label{fig: reg80}
\end{figure}

\begin{figure}[!h]
    \centering
    \includegraphics[width=0.99\linewidth]{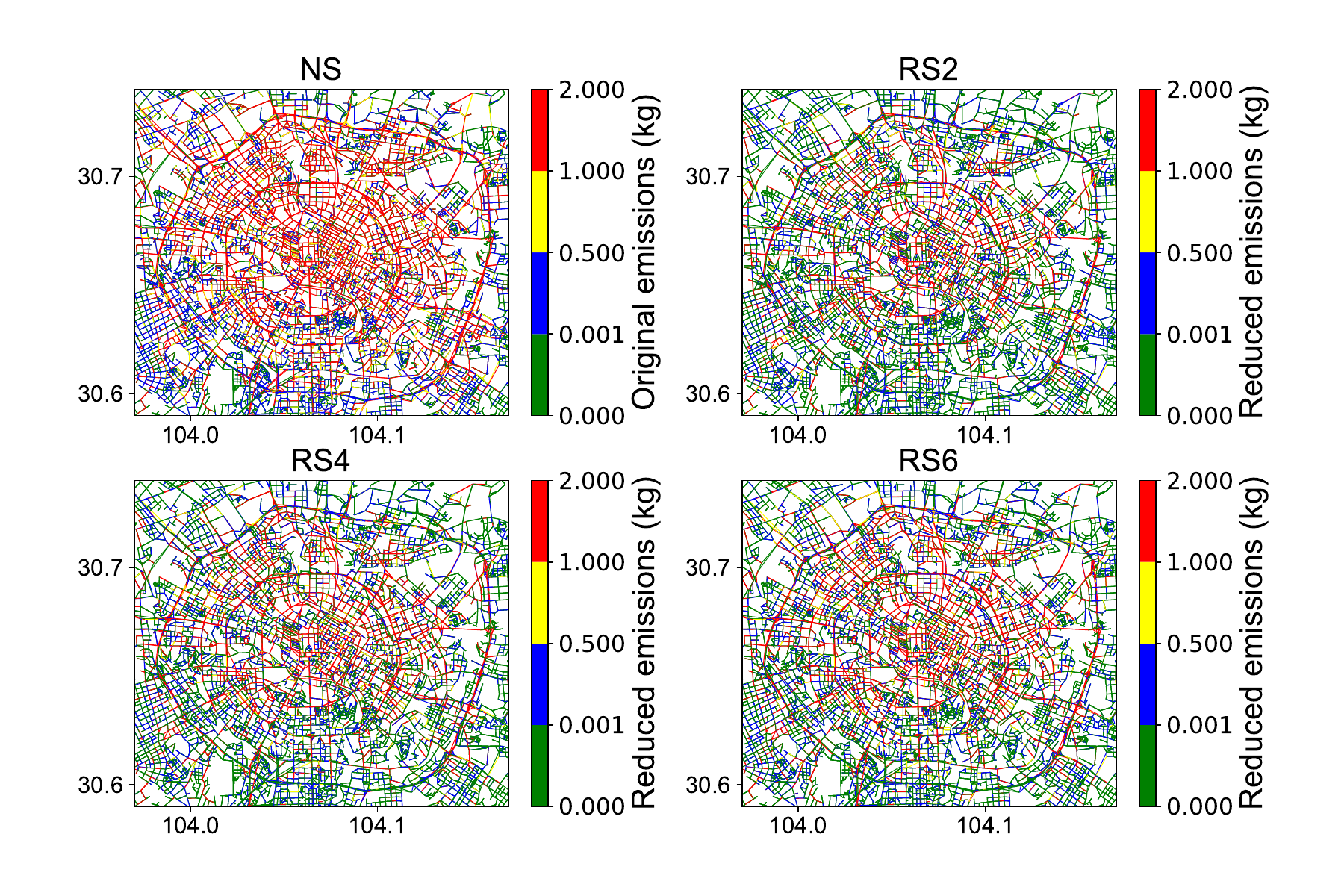}
    \caption{Spatial distributions of cumulative CO$_2$ emissions from non-sharing ride-sourcing and reductions from ride-sharing at 80\% of SR level.}
    \label{fig: vis80}
\end{figure}

\subsection{Marginal benefits of increase in vehicle capacity}\label{sec: 4.4}

Figure \ref{fig: result2} illustrates that, across all measurement metrics and various required service reliability (SR) levels, there are significant marginal benefits associated with increasing vehicle capacity when the capacity is no more than 4 passengers. However, when the vehicle capacity increases to 6 passengers, the marginal improvement in system efficiency, reductions in traffic emissions, and alleviation in traffic congestion resulting from higher-capacity ride-sharing services are not significant. For example, as shown in Fig. \ref{fig: result2} (c), when the SR is at 90\%, RS2, and RS4 can reduce CO$_2$ emissions by 31.7\% and 45.9\%, respectively. However, RS6 only decreases CO$_2$ emissions by 50.2\%, which is not significantly different from the reduction provided by RS4. This is because it is more challenging for multiple passengers to share a trip as more constraints need to be satisfied. Therefore, although the RS6 system allows for up to 6 passengers to share a vehicle on each trip, it is almost impossible to schedule 6 passengers with similar origins and destinations to a driver at one time.

\begin{figure}[!htbp]
    \centering
    \includegraphics[width=0.6\linewidth]{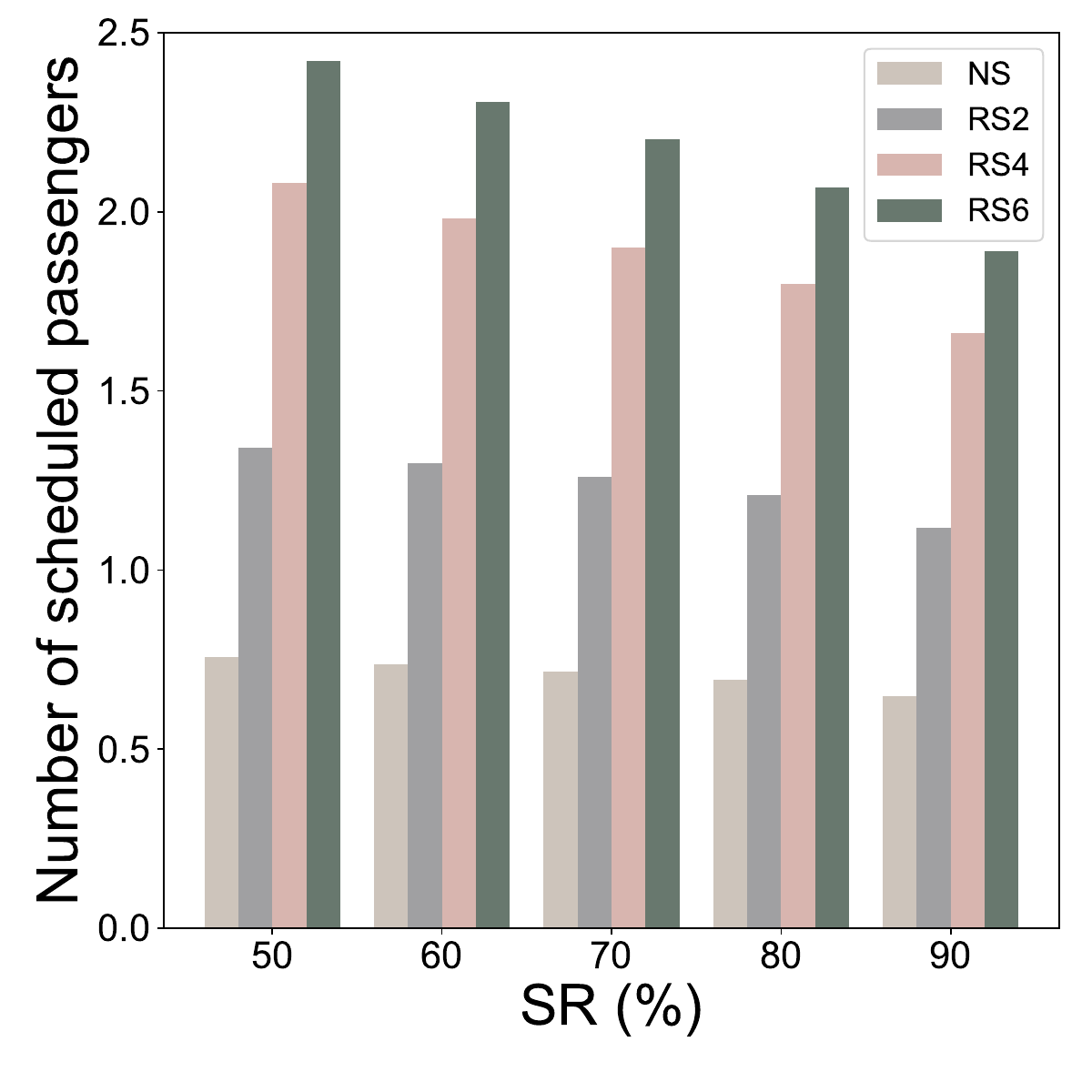}
    \caption{Average number of scheduled passengers.}
    \label{fig: NumPas}
\end{figure}

As shown in Fig. \ref{fig: NumPas}, this study calculates the average number of scheduled passengers of each vehicle, including the passengers already on the vehicle and those scheduled to be picked up in the further. The marginal improvement in the number of scheduled passengers with each increase in vehicle capacity is more significant in the scenarios where the capacity is no more than 4 passengers, compared to scenarios with a capacity of 6 passengers. For example, when the SR is at 80\%, RS2 and RS4 increase the number of scheduled passengers from 0.7 to 1.2 and 1.8 passengers, respectively, resulting in marginal improvements of 0.25 and 0.3 with one increase in vehicle capacity. In contrast, compared to RS4, RS6 only increases the number of scheduled passengers from 1.8 to 2.0, representing a marginal improvement of only 0.1. Moreover, in situations where the SR is low, indicating a market with a high demand-to-supply ratio where passengers find it easier to share their trips, the marginal increase in the number of scheduled passengers is more substantial. As a result, ride-sharing programs with no more than 4 passengers sharing a vehicle are generally recommended in practical ride-sourcing systems.

\section{Conclusion}\label{sec: 6}

This study develops an open-source ride-sharing simulation platform to investigate the impacts of high-capacity ride-sharing services on reducing traffic emissions and alleviating traffic congestion. A heuristic algorithm is developed to solve high-capacity ride-sharing problems, which significantly speeds up the simulation process while achieving high-accuracy results. Moreover, the platform integrates a speed-density traffic flow model and a typical traffic emission model to account for the interplay among traffic flow of regular and ride-sharing vehicles, traffic speed, and traffic emissions, enabling the platform to quantify traffic emission reductions and traffic congestion alleviation simultaneously. By conducting extensive experiments in Chengdu, this study finds that ride-sharing services with vehicle capacities of 2, 4, and 6 passengers can alleviate total traffic congestion by approximately 3\%, 4\%, and 5\%, respectively, and reduce traffic emissions from ride-sourcing systems by approximately 30\%, 45\%, and 50\%, respectively. In addition, this study finds that the carbon emission reduction attributed to the increased traffic speed after the implementation of ride-sharing only provides for a further 0.02-0.85\% of traffic emission reductions. Furthermore, we observe that the impacts of ride-sharing services on reducing traffic emissions and alleviating traffic congestion are more significant in hot areas with higher passenger demand. It is also interesting to find that the marginal benefits of increasing vehicle capacity in ride-sharing programs are insignificant when the vehicle capacity exceeds 4. These results provide valuable insights into managing, operating, and regulating on-demand ride services. For example, it is not recommended to implement high-capacity ride-sharing programs carrying more than 4 passengers under situations in which demand is not extremely high, since the additional benefits brought by the increasing vehicle capacity become less significant when the vehicle capacity is more than 4.

This study opens up a few potential avenues for future research. First, our work provides a foundation for the design of a reasonable price strategy to attract more passengers to share trips, thereby leading to a more efficient mobility system with significantly fewer traffic emissions. Second, car-following behavior models can also be integrated into our developed simulation platform to allow for microscopic simulations, which can more accurately quantify traffic emissions and congestion by simulating vehicles' acceleration and deceleration processes. Third, our developed simulation platform can be used to investigate the environmental impacts of future shared autonomous mobility systems in which all vehicles are connected, coordinated and fully controlled by the central platform for meeting mobility demands. Such research could provide valuable insights into the development of more sustainable, efficient, and equitable mobility systems in the near future.

\section*{Acknowledgement}
The corresponding author Jintao Ke would like to acknowledge the support from Hong Kong Environment and Conservation Fund (ECF) under Project No. 102/2022.

\newpage

\bibliography{reference}

\setcounter{equation}{0}
\setcounter{figure}{0}
\setcounter{table}{0}
\setcounter{section}{0}
\renewcommand\theequation{S.\arabic{equation}}
\renewcommand\thefigure{S\arabic{figure}}
\renewcommand\thetable{S\arabic{table}}
\renewcommand\thesection{S\arabic{section}}

\newpage
\section{Appendix}\label{sec: SI}

\begingroup
\setlength{\tabcolsep}{6pt} 
\renewcommand{\arraystretch}{1} 
\begin{table}[!htbp]
\caption{Comparison of the impacts of ride-sharing services on traffic emission reductions and traffic congestion alleviation in the hot zone and other areas at other SR levels.}
\begin{center}
\begin{tabular}{ c|c|c|c|c|c|c } 
 \hline
 SR@50\% & RS2 (other) & RS2(hot) & RS4 (other) & RS4(hot) & RS6 (other) & RS6(hot)\\
 \hline
CO$_2$ (\%) & 32.6 & 35.1 & 53.2 & 52.6 & 58.3 & 56.8 \\
CO (\%) & 32.6 & 35.4 & 53.0 & 52.6 & 58.4 & 56.9 \\
NOx (\%) & 32.4 & 35.2 & 52.9 & 52.5 & 58.3 & 56.9 \\
HC (\%) & 32.5 & 35.1 & 53.1 & 52.5 & 58.2 & 56.9 \\
DF (\%) & 2.4 & 2.6 & 3.5 & 3.8 & 3.8 & 4.1 \\
\hline
 SR@60\% & RS2 (other) & RS2(hot) & RS4 (other) & RS4(hot) & RS6 (other) & RS6(hot)\\
\hline
CO$_2$ (\%) & 31.2 & 35.1 & 50.9 & 51.8 & 55.7 & 56.5 \\
CO (\%) & 31.2 & 35.0 & 50.6 & 51.6 & 55.5 & 56.4 \\
NOx (\%) & 31.2 & 35.0 & 50.5 & 51.6 & 55.6 & 56.3 \\
HC (\%) & 31.0 & 35.0 & 50.6 & 51.6 & 55.8 & 56.5 \\
DF (\%) & 2.7 & 2.9 & 3.9 & 4.3 & 4.2 & 4.6 \\
\hline
 SR@70\% & RS2 (other) & RS2(hot) & RS4 (other) & RS4(hot) & RS6 (other) & RS6(hot)\\
\hline
CO$_2$ (\%) & 30.5 & 33.9 & 48.0 & 51.1 & 53.0 & 56.0 \\
CO (\%) & 30.6 & 33.9 & 48.1 & 51.1 & 53.2 & 55.8 \\
NOx (\%) & 30.5 & 34.0 & 48.1 & 51.1 & 53.0 & 55.8 \\
HC (\%) & 30.4 & 34.0 & 48.1 & 51.0 & 53.2 & 55.7 \\
DF (\%) & 2.9 & 3.2 & 4.3 & 4.7 & 4.6 & 5.1 \\
\hline
 SR@90\% & RS2 (other) & RS2(hot) & RS4 (other) & RS4(hot) & RS6 (other) & RS6(hot)\\
\hline
CO$_2$ (\%) & 30.1 & 31.2 & 45.0 & 47.8 & 49.0 & 52.3 \\
CO (\%) & 30.2 & 31.3 & 45.0 & 47.9 & 48.7 & 52.0 \\
NOx (\%) & 30.2 & 31.3 & 44.9 & 47.8 & 48.7 & 52.3 \\
HC (\%) & 30.3 & 31.4 & 44.7 & 47.9 & 48.8 & 52.3 \\
DF (\%) & 3.4 & 3.9 & 4.9 & 5.5 & 5.3 & 6.0 \\

\hline
\end{tabular}
\label{tableS1}
\end{center}
\end{table}

\begin{figure}
    \centering
    \subfigure[At 50\% of SR level]{\includegraphics[width=0.42\linewidth]{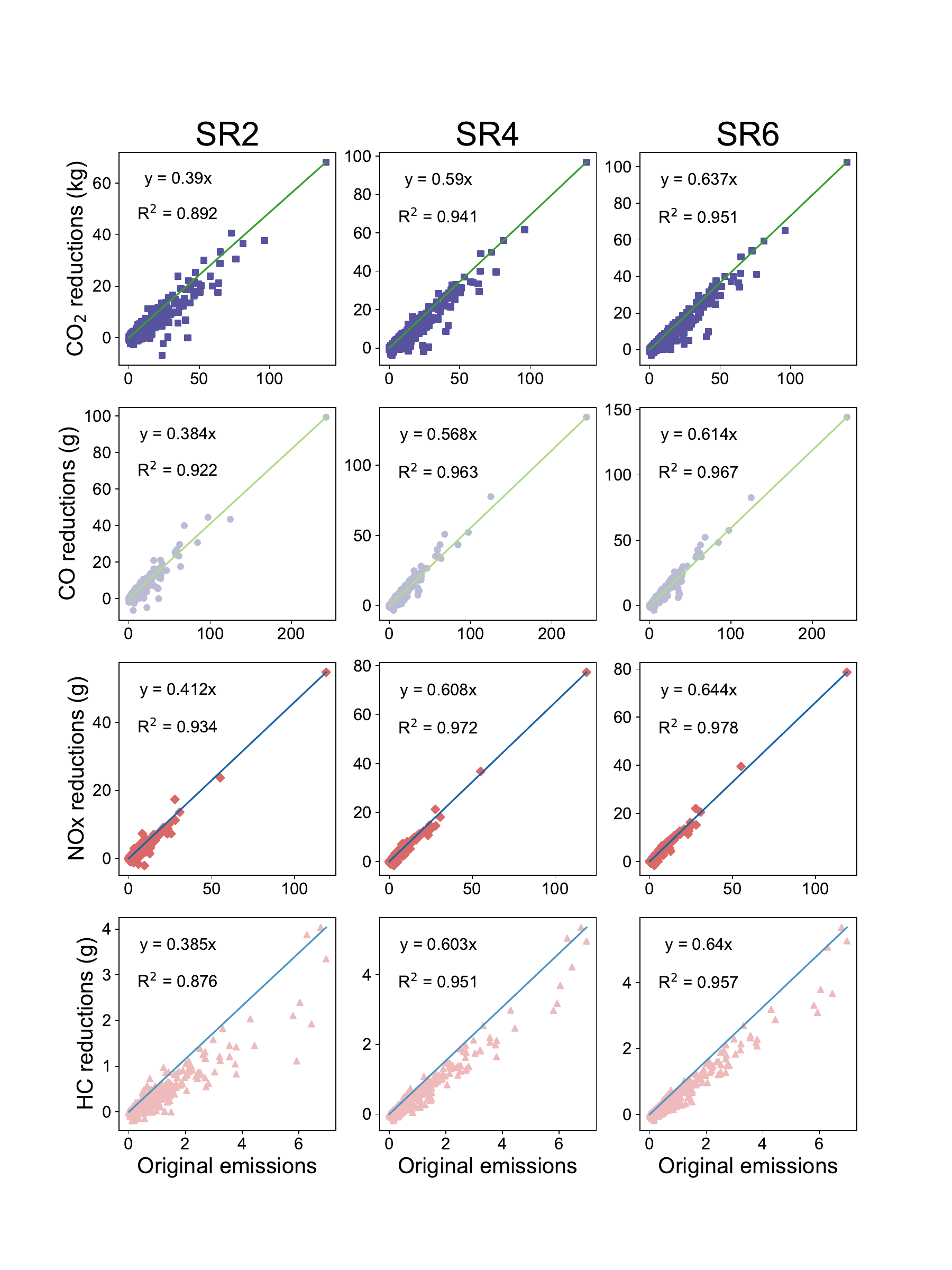}}
    \subfigure[At 60\% of SR level]{\includegraphics[width=0.42\linewidth]{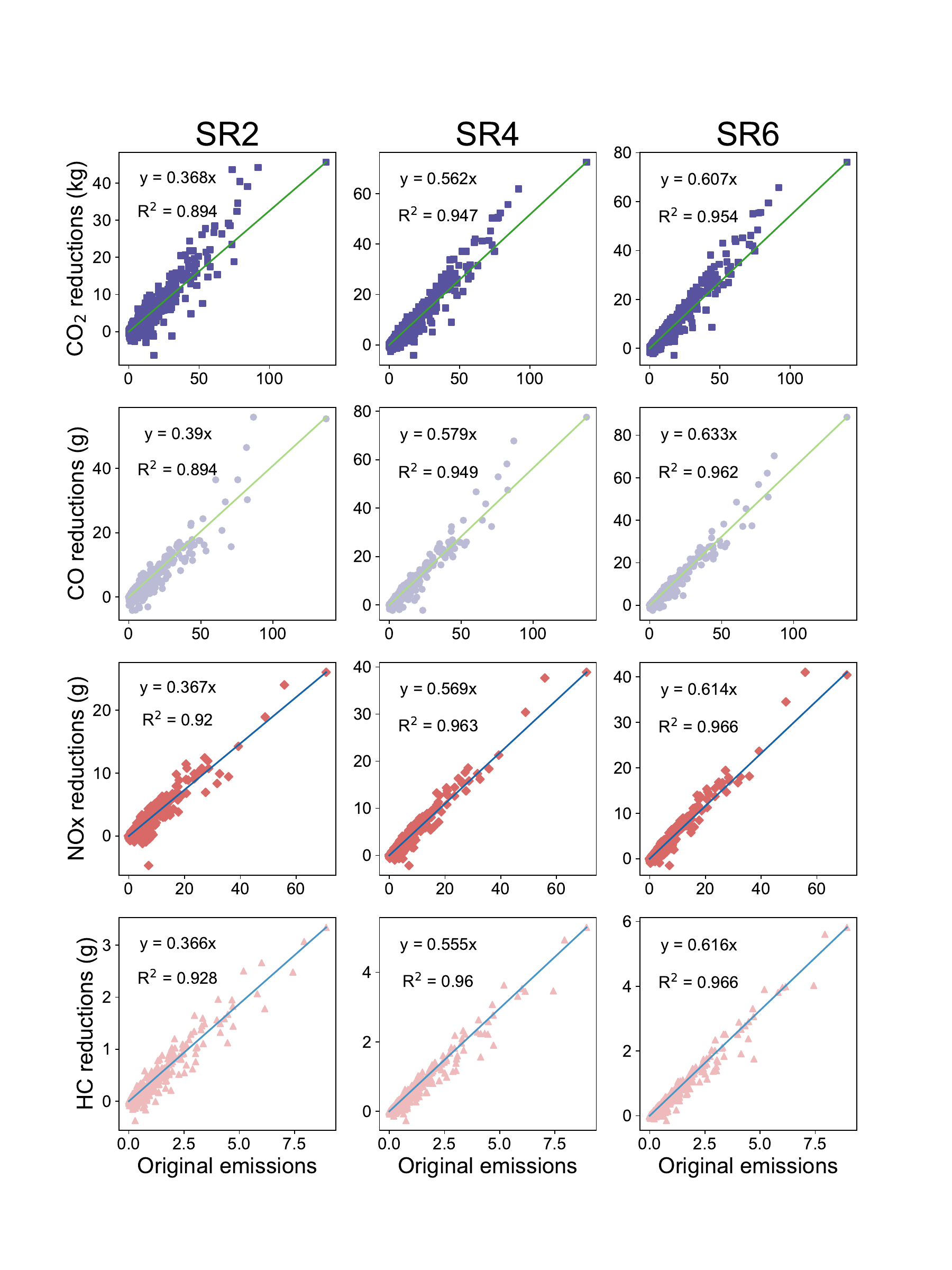}}
    \subfigure[At 70\% of SR level]{\includegraphics[width=0.42\linewidth]{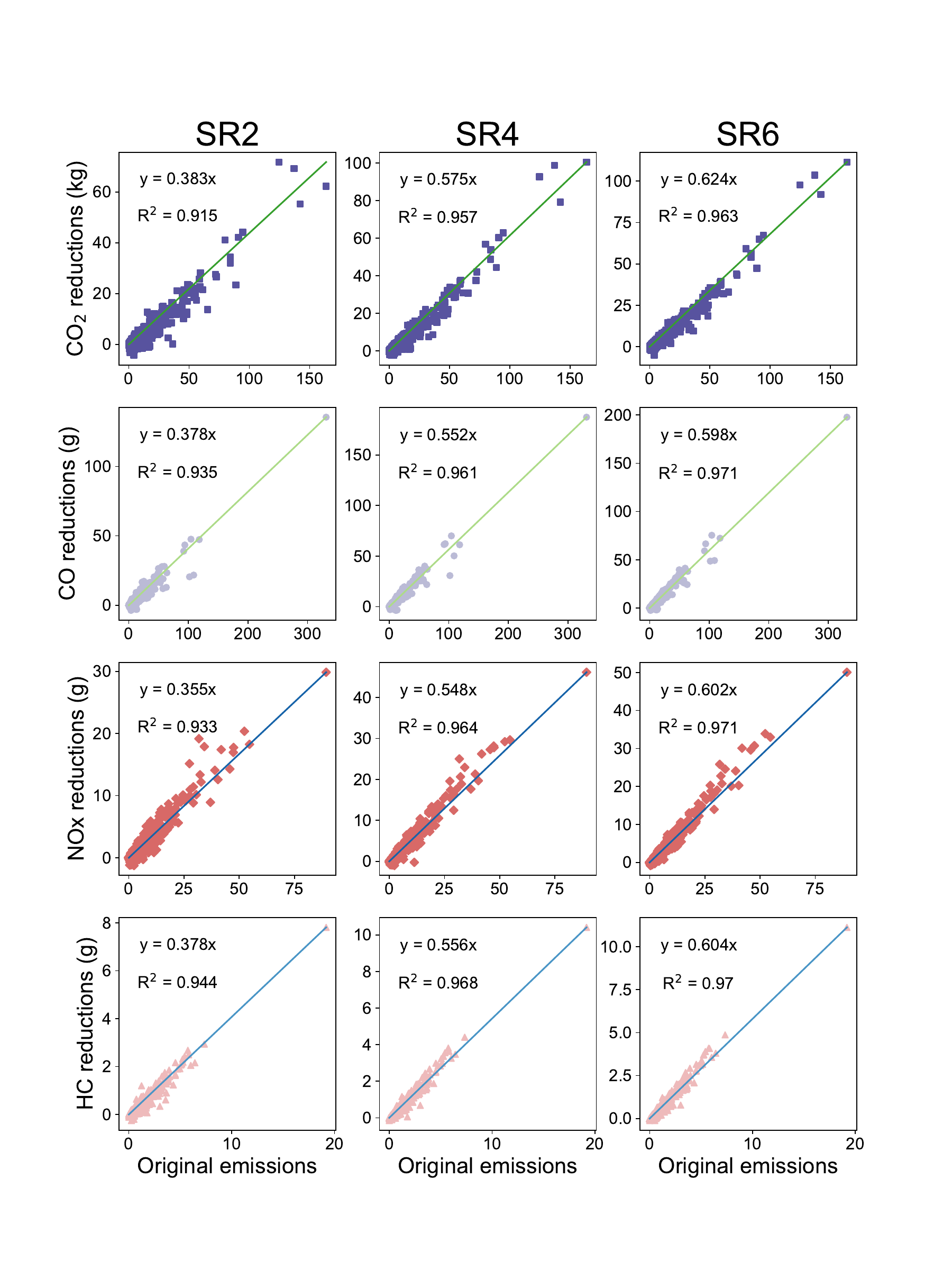}}
    \subfigure[At 90\% of SR level]{\includegraphics[width=0.42\linewidth]{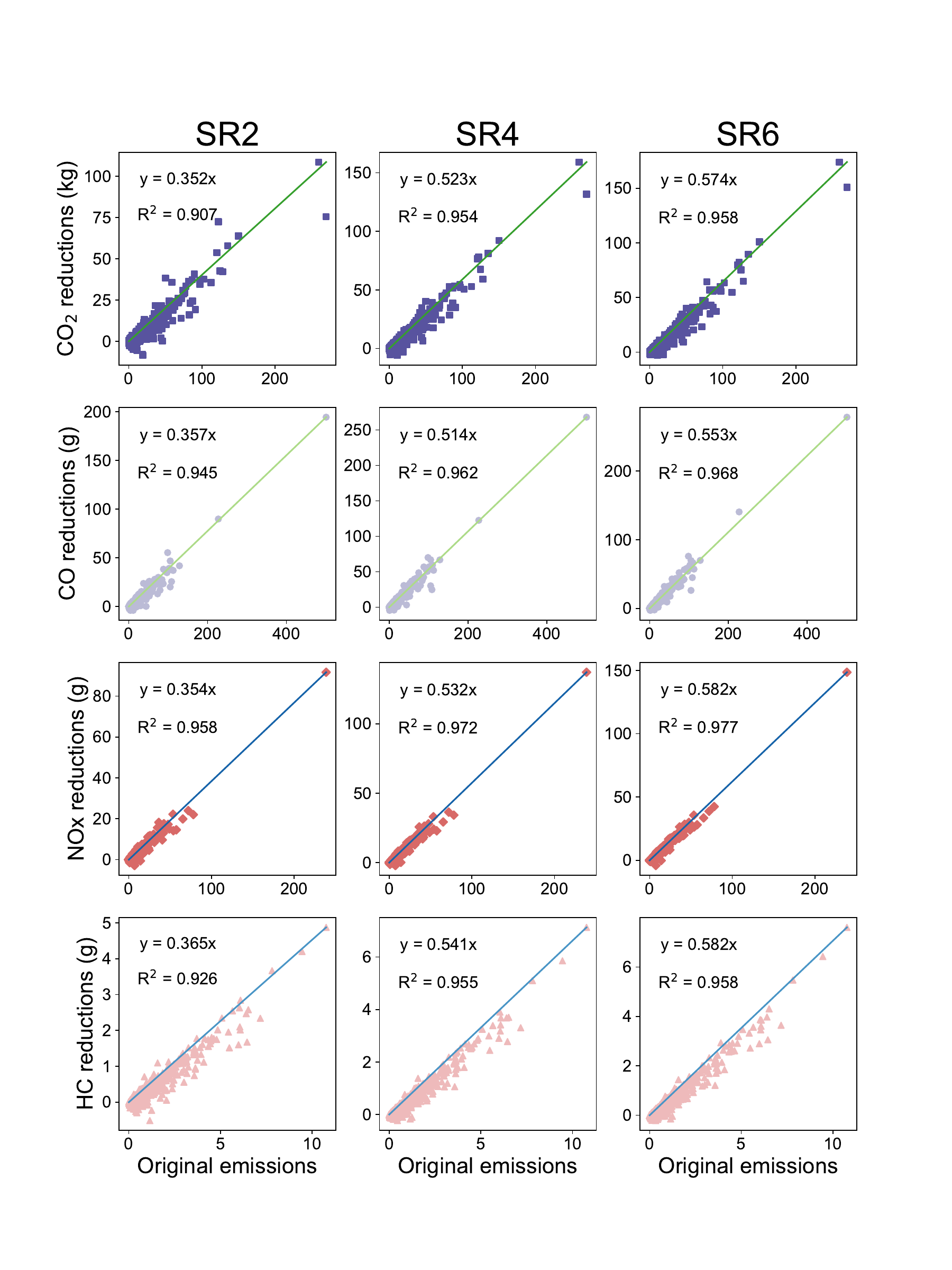}}
    \caption{Relationship between reduced traffic emissions brought by ride-sharing and original emissions from non-sharing ride-sourcing on each road at other SR levels.}
    \label{fig: S1}
\end{figure}

\end{document}